\documentstyle[11pt]{article}
\newcommand{\tr}{\mbox{Tr}}

\begin{document}

\thispagestyle{empty}
%\vspace*{1cm}

\begin{center}

{\Large Chiral Perturbation Theory:  a Primer}

\end{center}

\vfill

\begin{center}

 Barry R. Holstein

Department of Physics and Astronomy

University of Massachusetts

Amherst, MA  01003

and

Institute for Nuclear Theory, NK-12

University of Washington

Seattle, WA  98195

\end{center}

\vspace{1cm}

\begin{center}
{\bf Abstract}
\end{center}

Recently methods have been developed which exploit the chiral symmetry of QCD
in
order to make rigorous contact with low energy particle physics
phenomenology.  In
these lectures we present a pedagogical introduction to these techniques.

\vfill

\begin{center}

Research supported in part by the National Science Foundation and by the
Department of Energy

\end{center}

\newpage
\tableofcontents
\newpage
\section{Introduction}

For the past quarter century a primary goal of both particle and nuclear
theorists has been the ability to make reliable predictions about experimental
quantities directly from the Lagrangian of QCD.  However, this
quest has proven to be extraordinarily difficult because QCD, while
formally similar to QED, also possesses important differences.

\subsection{QED}

In Quantum Electrodynamics the interaction between charged particles is
mediated
by the exchange of {\it neutral} gauge bosons---photons.
Because of the neutrality of the photon there do {\it not }exist
vertices where
a photon interacts directly with another photon.  Therefore in QED only a
single vertex is required---{\it i.e.} the coupling of the photon to a fermion.
The QED Lagrangian density is
\begin{equation}
{\cal L}_{\mbox{{\tiny QED}}}= \bar q (i  {\not\!\! D} - m ) q -
{1 \over 4} F_{\mu\nu} F^{\mu\nu} \,  ,
\end {equation}
where
\begin{equation}
i D_\mu = i \partial_\mu - Q_qe A_\mu
\end{equation}
is the covariant derivative and
\begin{equation}
F_{\mu\nu}=\partial_\mu  A_\nu -  \partial_\nu  A_\mu  \, .
\end{equation}
is the electromagnetic field tensor.  The coupling constant e in QED is
related to the fine structure constant via $\alpha=e^2/4\pi
\approx 1/137$, and, because of the smallness of $\alpha$, the theory
can be successfully treated perturbatively.  Quantum Electrodynamics has
thereby
been confronted with numerous precise
experimental tests and has proven remarkably successful in each case.

\subsection{Quantum Chromodynamics}
The remarkable success of QED leads quite naturally to a nonabelian
generalization involving a triplet of color-charges interacting via the
exchange of color gauge bosons called gluons.  This is the theory of
Quantum Chromodynamics with the Lagrange density
\begin{equation}
{\cal L}_{\mbox{\tiny QCD}}=\bar{q}(i  {\not\!\! D} - m )q-
{1\over 2} \tr \; G_{\mu\nu}G^{\mu\nu} \, .
\end{equation}
Here the covariant derivative is
\begin{equation}
i D_{\mu}=i\partial_{\mu}-gA_\mu^a {\lambda^a \over 2} \, ,
\end{equation}
where $\lambda^a$ (with $a=1,\ldots,8$) are the SU(3) Gell-Mann matrices,
operating in color space.  The color-field tensor is defined by
\begin{equation}
G_{\mu\nu}=\partial_\mu  A_\nu -  \partial_\nu  A_\mu -
g [A_\mu,A_\nu]  \, ,
\end{equation}
\begin{samepage}
where the last term, {\it which has no QED analog}, arises from the
non-abelian nature of the theory.
Despite the {\it formal} similarity of the QCD and QED Lagrangians,
more careful examination
reveals crucial differences between the two theories:

\begin{itemize}
\item[--] i) The coupling constant $g^2/4\pi \sim 1$ so that a
perturbative treatment analogous to that used for QED is not possible.
\item[--] ii) In QCD gauge bosons themselves
carry color-charge.  Therefore
we have, in addition to the fermion--gluon vertex, three-
and four-gluon vertices,
which makes the theory highly nonlinear.
(A corresponding  situation exists in general relativity, where gravitons
themselves carry energy-momentum and therefore couple to one another).
\end{itemize}
\end{samepage}
These difficulties have heretofore prevented a precise confrontation of
experiment with
rigorous QCD predictions.  Nevertheless there are at least two cases in which
these
problems can be ameliorated and reliable theoretical predictions can be
generated
from QCD:
\begin{itemize}
\item[--] High energy limit:
At very high energies, when the momentum transfer
$q^2$ is large, QCD becomes ``asymptotically free"---{\it i.e.}
the running coupling constant $g(q^2)$ approaches
zero.  Hence, in this limit one can utilize perturbative methods.
However, this procedure, ``{\it perturbative QCD},"
is not useful except for interactions at the very highest energies.\cite{1}

\item[--] Symmetry:
The second way to confront QCD with experimental test is to utilize the
{\it symmetry} of
${\cal L}_{\mbox{{\tiny QCD}}}$. In order to do so, we separate the quark
components into two groups.  That involving the heavy quarks-- c,b,t---we
shall not consider further in these lectures.
Indeed the masses of such
quarks are much larger than the QCD scale---$\Lambda_{QCD}\sim 300$ MeV---but
can be treated using {\it heavy-quark symmetry} methods.  On the other hand,
the
light quarks---u,d,s---have masses much smaller than the QCD
scale and their interactions can be analyzed by exploiting the
{\it chiral symmetry} of the QCD Lagrangian as will be developed
further below.\cite{3}  As we shall see, this procedure is
capable of rigor but is only useful for energies $ E << 1$ GeV---it is a
{\it low energy method}.
\end{itemize}

However, before going into detail about chiral techniques it is useful to
review general ideas about symmetry and in particular about
symmetry {\it breaking.}

\section{Symmetry and Symmetry Breaking}
\subsection{Symmetry}
The best definition of symmetry for our purposes
is probably that due to the mathematician Herman Weyl who
said that a system is symmetric when one can do something to it and,
after making this change, the system looks the same as it did before.\cite{4}
The importance of symmetry in physics is due to an important
result---Noether's theorem---which connects
each symmetry of a system with a corresponding conserved current and
conservation law.\cite{5}  Familiar examples include:
\begin{itemize}
\item[--] ${\cal L} $ invariant under translation $\rightarrow$ momentum
conservation \par
\item[--] ${\cal L}$ invariant under time translation $\rightarrow$
energy conservation \par
\item[--] ${\cal L}$ invariant under rotation $\rightarrow$ angular
momentum conservation
\end{itemize}
However, because these symmetries are so familiar (and are exact) they are
also not of interest here.  Rather we shall be dealing in these lectures
with examples of
approximate symmetries which would obtain in some
hypothetical universe which is not our own---in our world
such symmetries will
be seen to be broken in some fashion.  In spite of this,
such broken symmetries are
of great
importance and by their study we will be able to learn much
about the underlying interactions.
\subsection{Symmetry Breaking} \par
In general there exist in physics three possible mechanisms for symmetry
breaking
\begin{itemize}
\item[--] explicit symmetry breaking
\item[--] spontaneous symmetry breaking
\item[--] quantum mechanical symmetry breaking
\end{itemize}
and in this section we study examples of each:

\subsubsection{Explicit Symmetry Breaking}
First consider a simple harmonic oscillator of frequency $\omega_0$
described by the Lagrangian

\begin{equation}
L= {1\over 2}m {\dot x}^2-{1\over 2} m \omega_0^2 x^2 \, .
\end{equation}
which is explicitly invariant under spatial
inversion---$x\rightarrow -x$---since
\begin{equation}
V_0(x)= -{1 \over 2} m\omega_0^2 x^2=-{1 \over 2} m\omega^2
(-x)^2=V_0(-x)
\end{equation}
Thus it is clear from symmetry considerations that the equilibrium
location $x_E$, which is determined
by the condition $[\partial L/ \partial x](x_E)=0$,
must occur at $x_E=0$, since the equilibrium position
should also manifest this symmetry.

Now, however, consider what happens if we add an term $V_1(x)=\lambda x$
{\it i.e.} a constant force, to the Lagrangian.   The new Lagrangian
is
\begin{equation}
L={1\over 2}m\dot{x}^2-{1\over 2}m\omega_0^2x^2+\lambda x
\end{equation}
which describes a displaced oscillator.  This new Lagrangian is {\it not}
invariant under spatial inversion, and consequently the new equilibrium
location---$x_E=\lambda / m\omega^2\neq 0$---is no longer required
to be at the origin.  This is an example of {\it explicit symmetry breaking}
wherein the symmetry violation is manifested in the Lagrangian itself.

\subsubsection{Spontaneous Symmetry Breaking}
As our second example, consider a hoop rotating in the earth's gravitational
field about a vertical axis.\cite{6}  Attached to the hoop is a bead which can
slide
along the circumference without friction.  The lagrangian $L$ for the system
is then
\begin{equation}
L= {1\over 2} m (R^2 {\dot  \theta}^2+\omega^2 R^2 \sin^2\theta)
+m g R \cos \theta \, ,
\end{equation}
where $\theta$ measures the angular displacement of the bead from the nadir.
$L $ is clearly symmetric under the angular parity transformation
$L(\theta)=L(-\theta)$, but the equilibrium condition for the bead
is found to be
\begin{equation}
{\partial L\over \partial \theta }=m \omega^2 R^2\sin\theta(\cos\theta-
{ g\over {\omega^2 R}})=0  \, .
\end{equation}
which is somewhat more complex than the displaced oscillator considered above.
For slow rotation---{\it i.e } for $\omega^2<{g\over R}$, we have
$\cos\theta-{g\over{\omega^2 R}} \neq 0$,
so that the ground (equilibrium) state configuration is given by $\theta_E=0$
as expected from symmetry considerations.  However, if we proceed to higher
angular velocities such that $\omega^2 > {g\over R}$ then
the bead finds equilibrium at $\theta_E=\pm\cos^{-1}{g\over \omega^2 R}$,
where the choice of + vs. - is {\it not} determined by the physics but
rather by the
history of motion of the system as the critical angular velocity was
reached.  Note that neither of these equilibrium positions exhibits
the symmetry of the underlying potential, which is invariant
under the exchange of $\theta$ and $-\theta$.  This is an example of
{\it spontaneous symmetry breaking}, wherein the
Lagrangian of a system possess a symmetry, but this symmetry is broken
by the ground (equilibrium) state of the system.

\subsubsection{Quantum Mechanical Symmetry Breaking}

The third type of symmetry breaking is the least familiar to most
physicists because it
has no classical analog.  It is called ``quantum mechanical"
or ``anomalous" symmetry breaking and occurs when the classical
Lagrangian of a system possesses a symmetry, but the symmetry broken
in the process of quantization.
As the simplest example and the only one (of which I am aware) that does not
involve quantum field theory---just quantum mechanics!---consider a
free particle, for which the stationary state Schr\"odinger equation is\cite{7}

\begin{equation}
-{1\over 2 m} \nabla^2 \psi= E \psi\equiv {k^2\over 2m}\psi  \, ,
\end{equation}
A partial wave solution in polar coordinates is
\begin{equation} \psi(\vec{r})={1\over r} \chi_k(r) P_l (\cos\theta) \, ,
\end{equation}
where $\chi_k(\vec{r})$ satisfies the radial Schr\"odinger equation

\begin{equation} \left(-{d^2\over d r^2}+ {l(l+1)\over r^2} + k^2\right)
\chi_k(r)=0\, .\end{equation}
Here the central piece in the above differential operator is the
well-known ``centrifugal potential."  By inspection the radial Schr\"odinger
equation is invariant under a ``scale transformation"

\begin{equation} r \rightarrow \lambda r \quad\quad k \rightarrow {1\over
\lambda }k\, .\end{equation}
This {\it scale invariance} has an important physical consequence, which can
be seen if we expand a plane wave solution in terms of incoming and outgoing
partial waves

\begin{equation} e^{ikz} \stackrel{r \rightarrow \infty}{\longrightarrow}
{1\over 2 ikr}
\sum_l (2l+1) P_l (\cos\theta)\left(e^{ikr} -e^{-i(kr-l\pi)} \right)\,
,\end{equation}
We observe that in each partial wave the incoming and outgoing component
of the wavefunction differ by the centrifugal phase shift $l\pi$.
This phase shift must be {\it independent} of energy via scale invariance.

If we place the free particle in a potential $V(\vec {r})$ then the
scale invariance is broken.  The corresponding wave function expanded in
partial waves then becomes
\begin{equation}
\psi^{(+)}(\vec r) \stackrel{ r \rightarrow \infty} {\longrightarrow}
{1\over 2 i k r} \sum_l (2l+1) P_l (\cos \theta)
(e^{i(kr+2\delta_l (k))}-e^{-i(kr-l\pi)} )
\end{equation}
Usually this is written as
\begin{equation}
\psi^{(+)}(\vec r)=e^{ikx}+{e^{ikr}\over r}f_k(\theta)
\end{equation}
where the scattering amplitude is defined by

\begin{equation}  f_k(\theta)= \sum_l(2l+1) {e^{2i\delta_l(k)}-1 \over 2ik}
P_l(\cos \theta)\, .\end{equation}
Of course, the phase shifts $\delta_l(k)$ of various angular momenta $l$ now
depend on energy, but this is to be expected since the scale invariance no
longer obtains.

One can generalize the scattering formalism to two dimensions, in which case
we obtain for the scattering wave function

\begin{equation}\psi^{(+)}(\vec r)  \stackrel{ r \rightarrow \infty}
{\longrightarrow}
e^{ikz} +{1\over \sqrt{r}} e^{i(kr+{\pi \over 4})} f_k(\theta)\end{equation}
and for the scattering amplitude

\begin{equation} f_k(\theta)=-i\sum^{\infty}_{m=-\infty} {e^{2i\delta_m(k)}-1
\over
\sqrt{2 \pi k} }e^{im\theta} \, \end{equation}
where we expand in terms of exponentials $e^{im\theta}$ rather than Legendre
polynomials.  What is {\it special} about two dimensions is that it
is possible to introduce
a {\it scale invariant} potential
\begin{equation} V(\vec r)=g\delta^2(\vec r)\end{equation}
The associated differential scattering cross section is found to be\cite{8}
\begin{equation}{d \sigma \over d \Omega} \propto {\pi \over 2 k} {1\over
(\ln {k\over \mu}^2)} \, .\end{equation}
which is somewhat of a surprise.  Indeed since the cross section is isotropic,
the scattering is pure $m=0$, corresponding to a phase shift

\begin{equation}  \cot \delta_0(k)={1\over \pi} \ln {k^2\over \mu^2}-{2\over
g}\, ,\end{equation}
which depends on $k$---scale invariance has been broken
as a result of quantization.  Although this should not be completely unexpected
(indeed while at the classical level non-zero impact parameter means
no scattering, in quantum mechanics this is not the case because of the
non-zero deBroglie wavelength), still the ``physics" of this result is not
completely clear.

\section{Examples of Symmetries}
In this section we study examples of symmetry and symmetry breaking found
within the Lagrangian of QCD and discuss ways in which these features are
manifested in the interactions of hadronic systems.
\subsection{Explicitly Broken Symmetry}
To begin, we assume a Lagrangian within only the u,d quark sectors
\begin{eqnarray}
{\cal L} & = & \bar{u}(i  {\not\!\! D} - m_{u})u+
\bar{d}(i  {\not\!\! D} - m _{d})d \nonumber \\
&\equiv& \bar{q}(i  {\not\!\! D} - m )q \, ,
\end{eqnarray}
where $q$ and $m$ are defined as
\begin{equation}q=\left(\begin{array}{c}
u\\ d
\end{array}\right) \,
\qquad
m=\left(\begin{array}{c c}
m_u & 0 \\  0 & m_d
\end{array}\right) \, . \end{equation}
In the limit $m_u=m_d$ this Lagrangian is unchanged after arbitrary
rotations
\begin{equation} \label{su2}
 q\rightarrow \exp(i {1\over 2}\vec{\tau}\cdot\vec{\alpha})q \, ,
\end{equation}
where $\vec \tau$ represents the Pauli Matrices---{\it i.e.} the u,d
quark Lagrangian is SU(2) flavor (isotopic-spin) invariant.

Now define the vector current density
\begin{equation} \vec{V}_{\mu}=\bar q \gamma_{\mu}{1\over 2}\vec{\tau }q\,
,\end{equation}
which is conserved for equal masses $m_u=m_d$---
\begin{equation} \partial^{\mu}\vec{V_{\mu}}=0\, .\end{equation}
Therefore the associated isospin charge, given by
\begin{equation} \vec{I}\equiv \int{d^3x\vec{V_0}(\vec{x},t)} \,
.\end{equation}
is time-independent---
\begin{equation} {d\over dt}\vec{I}=\int{d^3x{\partial\vec {V_0}\over \partial
t}}
=-\int {d^3x\vec{\nabla}\cdot\vec{V}}=-\int{\vec{dS}\cdot\vec{V}}=0 \, .
\end{equation}
where we have used Gauss' theorem and locality in making the last step.

These isotopic charge operators form an SU(2) algebra with
\begin{equation} [I_i,I_j]=i \varepsilon_{ijk} I_k \, , \end{equation}
Since these commutation relations are identical to those for ordinary spin
we know that the eigenstates, eigenvalues must be identical to those for
spin, {\it i.e.}
\begin{eqnarray}
\hat I^2 |I,I_z\rangle &=& I(I+1)|I,I_z\rangle \nonumber \\
\hat I_z |I,I_z\rangle &=& I_z |I,I_z\rangle \, .\nonumber
\end{eqnarray}
Now since the Lagrangian ${\cal L}$ is unchanged under an isospin
rotation, states which differ only by $I_z$ must have identical spin-parity
assignments and be degenerate, as seen in Nature.

Since a rotation in isospin space merely changes the orientation of
the axes, ${\cal L}_{\rm eff}$  is invariant, where ${\cal L}_{\rm eff}$
represents an
{\it effective} Lagrangian which describes the interaction in terms of
{\it experimental} degrees of freedom (hadrons) instead of
{\it fundamental} ones (quarks).  An example of such an effective Lagrangian
which describes the interactions of nucleons with pions is
\begin{equation}
{\cal L}_{{\mbox{{\tiny eff}}}}
(\pi NN)=g\bar{N}\gamma_5\vec{\tau}N\cdot\vec{\phi} \, .
\end{equation}
Of course, in the real world the masses of the light quarks are {\it unequal}
and isospin invariance is {\it broken}.

\[ {\cal L}_{\mbox{{\tiny QCD}}} \not\longrightarrow
{\cal L}_{\mbox{{\tiny QCD}}} \quad\hbox{if} \quad m_u\neq m_d. \]
However, the concept of isospin remains a useful one provided that the
breaking is not too large---{\it i.e.} provided that the u,d mass splitting is
small compared to $\Lambda_{\rm QCD}$.  In this case we can write the mass
matrix in the form
\begin{equation}
\left( \begin{array}{c c}
 m_u  & 0 \\ 0 & m_d
\end{array}\right)
= \hat{ m} \mbox{{\bf 1}} +{1\over 2}(m_u -m_d)\tau_3 \, ,
\end{equation}
with $ \hat{m }={1\over 2}(m_u+m_d)$ and can hope to treat the
isospin breaking ($\tau_3 $) part perturbatively.  For example, the
nucleon mass is given to first order in perturbation theory by
 \begin{eqnarray}
m_N&=&m_0\bar N N + m_1 \bar N \tau_3 N \nonumber \\
&=& \bar N \left( \begin{array}{c c}
m_0+m_1& 0 \\
0 & m_0 -m_1
\end{array}\right)N \, ,
\end{eqnarray}
where $m_0,m_1$ are unknown constants,
so that proton and neutron masses are no longer degenerate---
\begin{equation} m_n-m_p={m_1\over m_0}(m_p+m_n) .\end{equation}

The neutron and proton are distinguished in addition by
their charge, so isotopic spin invariance is also broken by
electromagnetism.  Hence in order to completely understand the n,p mass
difference we must also
include electromagnetic effects.  Now at the phenomenological level the nucleon
is a simple three quark object and its mass contains a contribution from the
Coulomb energy between quark pairs\cite{11}
\begin{eqnarray}
U_p&\sim& {\alpha \over <r>}\left({({2 \over 3})}^2+
2{ 2\over 3}\cdot -{1\over 3}\right)=0 \nonumber \\
U_n &\sim& {\alpha \over <r>}\left({(-{ 1\over 3})}^2+
2{ 2\over 3}\cdot -{1\over 3}\right)= -{\alpha \over 3<r>} \, ,
\end{eqnarray}
where $<r>$ represents some average radial quark separation distance
within the nucleon.
Then {\it e.g.} $m_p - m_n \sim {\alpha \over 3<r>} \sim 0.5$ MeV,
suggesting $m_d-m_u \sim 2$ MeV, but this is only a rough estimate.

Similar considerations apply if we extend our discussion to SU(3)
({\it i.e.} Gell-Mann's {\it Eightfold Way})\cite{12} by including the mass of
the
strange quark.  Since $m_s>>m_u,m_d$ the breaking effects would be expected to
be
somewhat larger, but SU(3) symmetry is still found to be a very useful concept.
We begin by defining a free Lagrange density for the three quark system
\begin{equation}
{\cal L} = \bar q (i  {\not\!\! D} - m)q \,   ,
\end{equation}
where $q$ and $m$ are now defined as
\begin{equation} q=\left( \begin{array}{c}
u\\ d \\ s
\end{array}\right) \,  ,\qquad
m=\left( \begin{array}{c c c }
m_u & 0 & 0 \\
0 & m_d& 0\\
 0 & 0 & m_s
\end{array}\right) \,  . \end{equation}
As in the SU(2) case, if $m_u=m_d=m_s$ this Lagrangian is manifestly
invariant under rotations in this {\it three} dimensional space
\begin{equation} \label{su3}
q\longrightarrow \exp (i{1\over 2} \sum_j \lambda_j \alpha_j) q \,  ,
\end{equation}
where $\lambda_j$ are the Gell--Mann matrices and an arbitrary rotation is
defined by the eight parameters $\alpha_j, j=1,2,\ldots 8$.

Analogous to the $2^2-1=3$ $\tau$ matrices, which satisfy
\begin{eqnarray}
[\tau_i,\tau_j]&=& 2i\epsilon_{ijk} \tau_k \nonumber \\
\{\tau_i,\tau_j\}&=&2\delta_{ij}  \,   .
\end{eqnarray}
the $3^2-1=8$  Gell-Mann matrices obey
\begin{eqnarray}
[\lambda_i,\lambda_j]&=&2i f_{ijk}\lambda_k \nonumber \\
\{\lambda_i,\lambda_j\}&=&2i d_{ijk}\lambda_k \,   .
\end{eqnarray}
Hence there exist {\it eight} vector currents
\begin{equation}
V^j_{\mu}={\bar q}\gamma_{\mu}{1\over 2}\lambda_j q  \,  ,
\end{equation}
which are conserved---
\begin{equation} \partial^\mu  V_{\mu}^j=0 \,   \end{equation}
and
\begin{equation}
0={d\over dt}\int d^3 x V_0^j(\vec{x},t)\equiv{d\over dt} F_j
\end{equation}
These SU(3) charge operators $F_j$ form an SU(3) algebra with
commutation relations
\begin{equation}
[F_i,F_j]=i f_{ijk}F_k  \,   .
\end{equation}

\bigskip

{\bf Representations of the SU(2) and SU(3) Group} \par \bigskip

A unitary representation of the SU(2) [SU(3)] group is a
mapping of the  $2\times 2$ matrices in Eq. 27
[the $3\times 3$ matrices in Eq. 40] onto unitary matrices
$D(U)$ which, in general, act in spaces with different dimensions.
In SU(2) the representations are well-known:
\begin{center}
\begin{tabular}{l l l l}
SU(2) & $\{1\}$ & $I_z=0$ & $\Lambda$, $\eta$ \\
      & $\{2\}$ & $I_z={1 \over 2},-{1 \over 2}$ &$(p,n) (K^+,K^-)$ \\
      & $\{3\}$ & $I_z=1,0,-1$  &$(\pi^+,\pi^0,\pi^-)
 (\Sigma^+,\Sigma^0,\Sigma^-)$ \\
      & $\{4\}$ & $I_z={3\over 2},{1\over 2},-{1\over 2},-{3\over 2}$
      & $(\Delta^{++}, \Delta^+,\Delta^0\Delta^-)$
\end{tabular}
\end{center}
Likewise the various representations of SU(3) can be worked out, with the
most important representation being the well-known octet.

In order to represent SU(3) invariant effective interactions it is useful to
define the $3 \times 3$ matrices
\begin{eqnarray}
{1 \over \sqrt{2}} \sum_j \lambda_j \phi_j =
\left( \begin{array}{c c c}
\frac{\pi^0}{\sqrt{2}}+\frac{\eta^0}{\sqrt{6}} &
 \pi^-  & K^- \\
\pi^+ & -{\pi^0 \over \sqrt{2}}+{\eta^0\over \sqrt{6}} &\bar K^0 \\
K^+  & K^0& -{2 \eta^0 \over \sqrt{6}}
\end{array}
\right) \equiv \Phi \nonumber\\
{1\over \sqrt{2}}\sum_j \lambda_j B_j =
\left( \begin{array}{c c c}
\frac{\Sigma^0}{\sqrt{2}} +\frac{\Lambda^0}{\sqrt{6}} &
\Sigma^-& \Xi^- \\
\Sigma^+ & -{\Sigma^0\over \sqrt{2}}+{\Lambda^0\over \sqrt{6}} &
\Xi^0\\
p & n & -{2\Lambda^0\over \sqrt{6}}
\end{array}\right)\equiv B
\end{eqnarray}
as the contraction of the SU(3) octet fields with the associated
Gell-Mann matrices.  Then the most general
SU(3) invariant interaction describing the interactions of the octet baryons
with the pseudoscalars can be written in terms of two arbitrary constants
$F,D$ as

\begin{equation} {\cal L}_{\mbox{eff}}(\Phi \bar{B} B)= D \tr \bar{B} \gamma_5
\{\Phi,B\}+F\tr\bar{B} \gamma_5[\Phi,B] \end{equation}
which is the SU(3) analog of the SU(2) relation

\begin{equation} {\cal L}_{\mbox{eff}}(\pi \bar{N} N)= g\bar{N} \gamma_5
\vec{\tau}N \cdot \vec{\phi}\end{equation}

The invariance of Eq. 49 is easily demonstrated since under an SU(3)
rotation

\begin{equation} U= \exp (i{1\over 2} \sum_j\lambda_j \alpha_j) \end{equation}
a matrix $M$ must transform as

\begin{equation} M\rightarrow U M U^{-1} \end{equation}
We see then that a structure such as

\begin{equation} \tr \bar{B} \Phi B \rightarrow \tr U\bar{B} U^{-1}
U \Phi U^{-1} U B U^{-1} = \tr \bar{B}\Phi B \end{equation}
is clearly invariant, as required.  According to Eq. 49 then arbitrary
$\bar{B}BP$ vertices can be expressed in terms just two constants
{\it e.g.}\cite{14}
\begin{eqnarray}
 g(\pi^+\bar{p} n) &= & F+ D \nonumber \\
g(K^+\bar{p}\Lambda) & = & -{1\over \sqrt{6}}
(D+3F)
\end{eqnarray}
and experimentally this prediction is found to work extremely well.
Having studied examples of explicit symmetry breaking via SU(2) and
SU(3) methods, we now move on to the case of spontaneous symmetry
breaking in QCD.
\subsection{Spontaneous Symmetry Breaking}

The classic example of spontaneous symmetry breaking is that of the
ferromagnet.  In this case one deals with a Hamiltonian of the form
\begin{equation} H \sim \lambda \sum_{i, j} \vec{\sigma}_i\cdot \vec{\sigma}_j
f_{ij} \end{equation}
which is clearly rotationally invariant.  Yet a permanent magnet selects a
definite direction in space along which it is magnetized---the ground
state does not share the symmetry of the underlying interaction.  Note
that just as in the case of the rotating hoop the direction selected by
the ground state is {\it not} a matter of physics but depends rather on the
history of the system.

\bigskip
{\bf Chiral Symmetry}\par\bigskip
In order to understand how spontaneous symmetry breaking occurs in QCD
we must introduce the idea of chirality, defined by the operators
\begin{equation} \Gamma_{L,R} = {1\over 2}(1\pm\gamma_5)={1\over 2}
\left( \begin{array}{c c }
1 & \mp 1 \\
\mp 1 & 1
\end{array}\right)
\end{equation}
which project left- and right-handed components of the Dirac wavefunction
via
\begin{equation} \psi_L = \Gamma_L \psi \qquad \psi_R=\Gamma_R
\psi \quad\mbox{with}\quad \psi=\psi_L+\psi_R \end{equation}
In terms of these chirality states the quark component of the QCD Lagrangian
can be written as
\begin{equation} \bar{q}(i\not\! \! D-m)q=\bar{q}_Li\not \! \! D q_L +
\bar{q}_Ri\not\!\! D q_R -\bar{q}_L m q_R-\bar{q}_R m
q_L \end{equation}
The reason that these chirality states are called left- and right-handed can
be seen by examining helicity eigenstates of the free Dirac equation.  In the
high energy (or massless) limit we have
\begin{equation}
 u(p)= \sqrt{{E+m\over 2E}}
\left( \begin{array}{c }
\chi \\  {\vec{\sigma}\cdot\vec{p} \over E+m}\chi
\end{array}\right)
\stackrel{E \gg m}{\sim} \sqrt{{1\over 2}}
\left( \begin{array}{c }
\chi \\  \vec{\sigma}\cdot\hat{p} \chi
\end{array}\right)
\end{equation}
Left- and right-handed helicity eigenstates then can be identified as
\begin{equation}
u_L(p)  \sim  \sqrt{1\over 2}
\left( \begin{array}{c}
\chi \\ -\chi
\end{array} \right),\qquad
u_R(p)  \sim  \sqrt{1\over 2}
\left( \begin{array}{c}
\chi \\ \chi
\end{array} \right)
\end{equation}
But note that
\begin{eqnarray}
 \Gamma_L u_L= u_L && \Gamma_R u_L=0 \nonumber \\
 \Gamma_R u_R= u_R && \Gamma_L u_R =0
\end{eqnarray}
{\it i.e.} in this limit chirality is identical with helicity.
\[ \Gamma_{L,R} \sim \mbox{helicity!} \]

With this background, we now return to QCD.  We observe that if $m=0$
then

\begin{equation} {\cal L}_{\rm QCD}=\bar{q}_L i \not\!\! D q_L +
\bar{q}_R i \not\!\! D q_R \end{equation}
would be invariant under independent global
left- and right-handed rotations
\begin{equation}
q_L  \rightarrow \exp (i \sum_j \lambda_j\alpha_j)
q_L,\qquad
q_R  \rightarrow \exp (i\sum_j \lambda_j \beta_j)
q_R
\end{equation}
(Of course, in this limit the heavy quark component is also invariant, but
since $m_{c,b,t} >> \Lambda_{\rm QCD}$ it would be silly to consider this as
even an approximate symmetry in the real world.)  This invariance is called
$SU(3)_L \bigotimes SU(3)_R$ or chiral $SU(3)\times SU(3)$.  Continuing
to neglect the light quark masses,
we see that in a chiral symmetric world one would have sixteen---eight
left-handed
and eight right-handed---conserved currents
\begin{equation} \bar{q}_L\gamma_{\mu} {1\over 2} \lambda_i q_L \, ,
\qquad \bar{q}_R\gamma_{\mu}{1\over 2}\lambda_i
q_R \end{equation}
Equivalently, by taking the sum and difference we would have eight vector and
eight axial vector conserved currents
\begin{equation}
V^i_{\mu}=\bar{q}\gamma_{\mu} {1\over 2}
\lambda_i q,\qquad
A^i_{\mu}=\bar{q}\gamma_{\mu}\gamma_5
 {1\over 2} \lambda_i q
\end{equation}
In the vector case, we have already examined the consequences.  There exist
eight time-independent generators
\begin{equation} F_i=\int d^3 x V^i_0(\vec{x},t) \end{equation}
and there exist supermultiplets of particles in the
configurations demanded by SU(3).

If chiral symmetry were realized in the conventional fashion one would
expect there also to exist corresponding nearly degenerate
opposite parity states generated
by the action of the time-independent axial charges
$F^{\sigma}_i= \int d^3 xA^i_0(\vec{x},t)$
on these states.  Indeed since
\begin{eqnarray}
H|P\rangle &= & E_P|P\rangle \nonumber \\
H(Q_5|P\rangle)&=&Q_5(H|P\rangle)
=  E_P(Q_5|P\rangle)
\end{eqnarray}
we see that $Q_5|P\rangle$ must also be an eigenstate of the Hamiltonian
with the same eigenvalue as $|P>$, which would seem to require the existence of
parity
doublets.  However, experimentally this does not appear to be the
case.
\bigskip
\subsection{Goldstone's Theorem}

One can resolve this apparent problem by postulating that the theorem is
avoided because the axial symmetry is
{\it spontaneously broken.}  Then according to a theorem due to Goldstone,
when a continuous symmetry is broken in this fashion there must also be
generated a massless boson having the quantum
numbers of the broken generator, in this case a pseudoscalar, and when the
axial charge acts on a single particle eigenstate one does not get a single
particle eigenstate of opposite parity in return.\cite{15} Rather one generates
one or more of these massless pseudoscalar bosons

\begin{equation} Q_5|P\rangle \sim |Pa \rangle +\cdots \end{equation}
This phenomenon is a well-known one in ferromagnetism, where, since it does
not cost any energy to rotate the spin direction, one can find correlated
groups of spins which develop in a wavelike fashion---a spin wave with
\begin{equation} E\sim {c\over \lambda} \sim \mbox{cp} \end{equation}
which is the dispersion formula associated with the existence of a massless
excitation.

A simple example can be studied within the context of scalar field theory.
The spin-zero Lagrangian is
\begin{equation} {\cal L} = {1\over 2} \partial_{\mu} \phi \partial^{\mu}
\phi -{1\over 2} m^2 \phi^2 \end{equation}
which can be verified since by applying the Euler-Lagrange equation
\begin{equation} 0= \partial^{\mu} {\delta {\cal L} \over \delta (
\partial^{\mu}\phi)} - {\delta {\cal L} \over \delta \phi}
\end{equation}
we obtain the Klein-Gordon equation
\begin{equation} \partial^{\mu}\partial_{\mu}\phi +m^2\phi =
(\Box +m^2) \phi =0 \end{equation}
Now, however, consider a {\it complex} field $\phi(x)$
which has the Lagrangian
\begin{equation} {\cal L} = |\partial_{\mu}\phi|^2 - V(|\phi|) \end{equation}
If we write things in terms of the modulus and phase of the field $\phi (x)$

\begin{equation}
 \phi = {1\over \sqrt{2}} \rho e^{i\theta},\qquad
 \partial_{\mu} \phi =  {1\over \sqrt{2} } e^{i\theta}
(\partial_{\mu}\rho+i\rho \partial_{\mu} \theta)
\end{equation}
then the potential depends only on the modulus $\rho (x)$ and therefore has
the shape shown of a Mexican hat, for which each point along the minimum
corresponds
to a different value for the phase but has the same energy.  The ground state
of course selects one particular value for the phase and breaks the
rotational symmetry.  Let this value be $\rho(x)=\rho_0, \theta =0$
and expand about this point $\rho =\rho^0 +\chi.$
The Lagrangian then reduces to
\begin{equation} {\cal L} ={1\over 2}(\partial_{\mu}\chi)^2+{1\over 2}
\rho^2_0(\partial_{\mu}\theta)^2-V({\rho^0\over\sqrt{2}})
-{1\over 2f}\chi^2 V'' ({\rho^0\over \sqrt{2}})+ \cdots \end{equation}

{\it i.e.}

\begin{equation} m^2_{\chi} = {1\over 2} V''({\rho^0 \over
\sqrt{2}}) \quad  m^2_{\theta}=0 \end{equation}
We see that there exists a massless excitation in the $\theta$
direction---this is the Goldstone mode.

According to this argument then one would expect there to exist eight
massless pseudoscalar states, which are the Goldstone bosons of QCD.
Examination of the particle data tables reveals that no such particles
exist, however, and causes us to ask what has gone wrong.  The answer
is found in the fact that our discussion thus far has neglected the piece
of the QCD Lagrangian which is associated with quark mass and can be
written in the form
\begin{equation} {\cal L}_{\mbox{QCD}}^m=-(\bar{u}_L u_R+\bar{u}_R
u_L) m_u-(\bar{d}_L d_R+\bar{d}_R d_L)m_d \end{equation}
Since clearly this term breaks the chiral symmetry---
\begin{eqnarray}
\bar{q}_L q_R &\rightarrow & \bar{q}_L \exp(-i
\sum_j \lambda_j\alpha_j) \times \exp(i\sum_j \lambda_j
\beta_j) q_R \nonumber \\
& \neq & \bar{q}_L q_R
\end{eqnarray}
---we have a violation of Goldstone's theorem.  The associated pseudoscalar
bosons are not required to be massless
\begin{equation} m^2_G \neq 0 \end{equation}
but since their mass arises only from the breaking of the symmetry the
various would-be Goldstone boson masses are expected to be proportional to
the breaking
\[ m^2_G \propto m_u,m_d,m_s \]
and therefore small to the extent that the quark masses are light. Indeed
the pseudoscalar masses are considerably lighter than other hadronic masses
in the spectrum, as expected in this scenario.

%\begin{figure}
%\end{figure}

\section{Effective Field Theory}

Before proceeding further it is useful to discuss the concept of effective
field theory, since it in this fashion that we will be able to
make contact with QCD.
An effective field theory is one which does not include all of the degrees
of freedom of the underlying (true) field theory and for this reason the
term {\it effective} is sometimes taken to mean {\it defective}.  However, this
is not at all necessarily the case, as the following example will show.
\subsection{Superconductivity}

An example which is somewhat close to the case of QCD is
that of superconductivity.  In this case the full degrees of freedom consist
of free electrons and a lattice of ions.  As is well known, the interaction
of one of these electrons with the lattice deforms the latter, which in turn
has an effect on a nearby electron, giving an effective binding between these
electron pair states.  In fact by integrating out the lattice completely one
has an effective field theory expressed entirely in terms of electron pair
states, which has the form\cite{17}

\begin{equation} L_{\mbox{eff}}= \chi^{\ast}\left( -{(
\vec{\nabla}+i e^{\ast}\vec{A})^2 \over 2m^{\ast}}
+e^{\ast}\phi\right) \chi a(T) +(\chi^{\ast}\chi)^2 b(T)
+\chi^{\ast}\chi c(T)
\end{equation}
with
\begin{equation} e^{\ast}=2e \qquad\qquad m^{\ast}=2m \end{equation}
The important feature here is the coefficient
\begin{equation} c(T)= K \ln {T\over T_c}\end{equation}
which changes sign as a function of temperature.  At temperatures $T>T_c$
the effective potential has the shape of a simple well {\it cf.} Figure  1a.
The ground state
occurs at $\chi^*\chi=0$ which means that there is nothing remarkable going
on.  On the other hand for $T<T_c$ this linear term changes sign so that
the effective potential now has the familiar double well behavior associated
with spontaneous symmetry breaking {\it cf.} Figure  1b.  The ground state
now occurs at $\chi^*\chi\ne 0$ which means that there occurs a Bose
condensation---the electron pairs condense into the same state and the
superconducting phase occurs.

\begin{figure}
\vspace{2.5in}
\caption{Shape of the effective superconducting potential above and below
the critical temperature.}
\end{figure}

The connection with the QCD problem can now be made.  Of course, instead
of weakly bound electron pairs interacting weakly with a lattice (which is
integrated
out of the effective Lagrangian) we have quark-antiquark pairs interacting
strongly with color gluons (which are integrated out of the effective
Lagrangian).  However, the idea is the same---in both cases we end up
with a description of the physics in terms of an effective interaction
which, even though not including all the relevant degrees of freedom,
nevertheless simply encapsulates the relevant physics in terms of those
degrees of freedom which are relevant experimentally.

\bigskip
%\begin{figure}
%\end{figure}

\begin{tabular}{l l l }
weakly bound  & $\rightarrow$ & strongly bound  \\
$L_{\mbox{eff}}$ for $(e^-e^-)$ & $\rightarrow$ &
$L^{\mbox{QCD}}_{\mbox{eff}}$ for $(q\bar{q})$ \\
%wavefunction $\chi$ & & wavefunction $\phi$ \\
Lattice degrees of freedom gone &$\rightarrow$ & Gluon
degrees of freedom gone \\
\end{tabular}
\bigskip

\subsection{Effective Chiral Lagrangian}
Our goal then is to generate an effective field theory which describes the
interactions
of the pseudoscalar (Goldstone) bosons by exhibiting the chiral symmetry
of QCD which
we have previously discussed.  This is done by defining a nonlinear function
of the pseudoscalar fields $U=\exp(i\vec{\tau}\cdot\pi/v$ such that under
the chiral transformations
\begin{eqnarray}
\psi_L & \rightarrow & L \psi_L \nonumber \\
\psi_R & \rightarrow & R \psi_R
\end{eqnarray}
then
\begin{equation} U \rightarrow L U R ^{\dagger} \end{equation}
and a form such as
\begin{equation} \mbox{Tr} \partial^{\mu} U \partial_{\mu} U^{\dagger}
\rightarrow \mbox{Tr} L \partial^{\mu} U R^{\dagger} R
\partial_{\mu} U^{\dagger} L^{\dagger} = \mbox{Tr} \partial^{\mu}
U \partial_{\mu} U^{\dagger}\,   \end{equation}
is invariant under chiral rotations and can be used as part of the
effective Lagrangian.  However, this form is
also not one which we can use in order to realistically describe
Goldstone interactions in Nature since according to Goldstone's theorem
a completely invariant Lagrangian must also have zero pion mass, in
contradiction to experiment.

We infer then that the {\it lowest order} effective chiral Lagrangian is
given by
\begin{equation}
 {\cal L}_2={v^2 \over 4} \mbox{Tr} (\partial_{\mu}U \partial^{\mu}
 U^{\dagger})+{m^2_{\pi}\over 4} v^2 \mbox{Tr} (U+U^{\dagger})\,  .
\end{equation}
where the subscript 2 indicates that we are working at two-derivative order
or one power of chiral symmetry breaking---{\it i.e.} $m_\pi^2$.
This Lagrangian is also {\it unique}---if we expand to lowest order in
$\vec\phi$
\begin{equation}
\mbox{Tr}\partial_{\mu} U \partial^{\mu} U^{\dagger} =
\mbox{Tr} {i\over v} \vec{\tau}\cdot\partial_{\mu}\vec{\phi} \times
{-i\over v}\vec{\tau}\cdot\partial^{\mu}\vec{\phi}= {2\over v^2}
\partial_{\mu}\vec{\phi}\cdot \partial^{\mu}\vec{\phi}\,  ,
\end{equation}
we reproduce the free pion Lagrangian, as required,
\begin{equation}
 {\cal L}_2 ={1\over 2} \partial_{\mu}
\vec{\phi}\cdot \partial^{\mu} \vec{\phi} -{1\over 2} m^2_{\pi}
\vec{\phi}\cdot \vec{\phi} +{\cal O} (\phi^4) \,  .
\end{equation}

At the SU(3) level, including a generalized chiral symmetry breaking term,
there is even predictive power---one has
\begin{equation}
 {v^2\over 4} \mbox{Tr} \partial_{\mu} U \partial^{\mu} U^{\dagger}
=   {1\over 2} \sum_{j=1}^8 \partial_{\mu}
\phi_j\partial^{\mu}\phi_j +\cdots \nonumber\\
\end{equation}
\begin{eqnarray}
{v^2 \over 4} \mbox{Tr} 2 B_0 m ( U+ U^{\dagger})
& =&  \mbox{const.}
-{1\over 2} (m_u+ m_d)B_0 \sum_{j=1}^3 \phi^2_j \nonumber\\
&-&{1\over 2} (\hat{m} +m_s)B_0\sum_{j=1}^3 \phi^2_j
 -{1\over 6} (m_u+m_d +4m_s)B_0\phi^2_8  +\cdots \, \nonumber\\
&&
\end{eqnarray}
where $B_0$ is a constant and $m$ is the quark mass matrix. We can
then identify the meson masses as
\begin{eqnarray}
 m^2_{\pi} & =& (m_u +m_d) B_0 = 2\hat{m} B_0
\nonumber \\
 m_K^2 &=& (\hat{m} +m_s) B_0 \nonumber \\
m_{\eta}^2 & =& {1\over 3} (m_u + m_d + 4 m_s) B_0 =
{2\over 3} (\hat{m} + 2m_s) B_0  \, ,
\end{eqnarray}
This system of three equations is {\it overdetermined}, and we find by simple
algebra
\begin{equation}
3m_{\eta}^2 +m_{\pi}^2 - 4m_K^2 =0 \, \, .
\end{equation}
which is the Gell-Mann-Okubo mass relation and is well-satisfied
experimentally.\cite{20}

\bigskip

{\bf Currents }\par\bigskip

In order to proceed further in our analysis, we can identify the Noether
currents via standard techniques.  Suppose that the
Lagrangian is invariant under the transformation
$\phi\longrightarrow \phi+\varepsilon f(\phi)$---{\it i.e.}
\begin{eqnarray}
0 & =& {\cal L} (\phi + \varepsilon f, \partial_{\mu} \phi +\varepsilon
\partial_{\mu} f) -{\cal L}(\phi, \partial_{\mu}\phi) \nonumber \\
&=&  \varepsilon f {\delta {\cal L} \over \delta \phi}
+\varepsilon \partial_{\mu} f {\delta {\cal L}\over \delta (
\partial_{\mu}\phi)}
=\varepsilon \partial_{\mu} \left( f {\delta {\cal L} \over
\delta(\partial_{\mu}\phi)}\right)  \, .
\end{eqnarray}
so that we can identify the associated conserved current as \footnote{Note:
This is often written in an alternative fashion by introducing a
{\it local} transformation
 $\varepsilon = \varepsilon(x)$, so that the Lagrangian transforms as
\begin{equation}
 {\cal L}(\phi,\partial_{\mu}\phi) \rightarrow
{\cal L} (\phi+\varepsilon f, \partial_{\mu} \phi +
\varepsilon \partial_{\mu} f +f \partial_{\mu}
\varepsilon) \, .
\end{equation}
Then
\begin{equation}
 {\partial {\cal L} \over \partial(\partial_{\mu}
\varepsilon)} = f {\partial {\cal L} \over \partial
(\partial_{\mu}\phi)} \,  ,
\end{equation}
so that the Noether current can also be written as
\begin{equation}
 J^{\mu} = {\partial {\cal L} \over \partial(
\partial_{\mu}\varepsilon)} \,  .
\end{equation}}
\begin{equation}
 J^{\mu} = f {\partial{\cal L} \over
 \partial(\partial_{\mu}\phi)} \, ,
\end{equation}
Since under a
\begin{equation}
\mbox{Vector, Axial transformation:}  \alpha_L  =\pm\alpha_R
\end{equation}
we have
\begin{equation}
U \rightarrow  \mbox{LUR}^{\dagger}  \stackrel{V}
{\simeq}  U + i\left[\sum_j \alpha_j\lambda_j, U
\right]
\stackrel{A}
{\simeq} U + i\left\{\sum_j \alpha_j\lambda_j, U
\right\}  \,  .
\end{equation}
which leads to the vector and axial-vector currents
\begin{equation}
\{V,A\}^k_{\mu}=  -i{v^2\over 4}\mbox{Tr}
\lambda^k(U^{\dagger}\partial_{\mu}U\pm U
\partial_{\mu}U^{\dagger})
\end{equation}

At this point the constant $v$ can be identified by using the axial current.
In SU(2) we find
\begin{eqnarray}
U^{\dagger}\partial_{\mu}U-U\partial_{\mu}
U^{\dagger} &= & 2i{1\over v} \vec{\tau}
\cdot\partial_{\mu}\vec{\phi}+ \cdots
\end{eqnarray}
so that
\begin{equation}
A^k_{\mu}=  i{v^2\over 4} \mbox{Tr}\tau^k
2i{1\over v} \vec{\tau}\cdot\partial_{\mu}\vec{\phi}
+\cdots = -v \partial_{\mu}\phi^k+\cdots \, .
\end{equation}
If we set $k=1-i2$ then this represents the axial-vector component of the
$\Delta S=0$ charged weak current and
\begin{equation}
 A_{\mu}^{1-i2}= -v\partial_{\mu}\phi^{1-i2}
=-\sqrt{2}v \partial_{\mu}\phi^- \,  .
\end{equation}
Comparing with the conventional definition
\begin{equation}
\langle 0|A_{\mu}^{1-i2}(0)|\pi^+
(p)\rangle = i\sqrt{2}F_{\pi}p_{\mu} \, ,
\end{equation}
we find that, to lowest order in chiral symmetry, $v=F_\pi$, where
$F_\pi=92.4$ MeV is the pion decay constant.\cite{21}

Likewise in SU(2), we note that
\begin{equation} U^{\dagger} \partial_{\mu} U + U\partial_{\mu}
U^{\dagger}= {2i\over v^2}\vec{\tau}\cdot\vec{ \phi}
\times \partial_{\mu}\vec{\phi}+\cdots \, , \end{equation}
so that the {\it vector current} is
\begin{eqnarray}
V_{\mu}^k & = & -i {v^2\over 4}\mbox{Tr}\tau^k
{2i\over v^2}\vec{\tau}\vec{\phi}\times
\partial_{\mu}\vec{\phi}+\cdots \nonumber \\
& =& (\vec{\phi}\times\partial_{\mu}\vec{\phi})^k+\cdot \, .
\end{eqnarray}
We can identify $V_\mu^k$ as the electromagnetic current by setting $k=3$
so that
\begin{equation}
 V_{\mu}^{\rm em}=\phi^+\partial_{\mu}\phi^-
-\phi^-\partial_{\mu}\phi^+ +\cdots
\end{equation}
Comparison with the conventional definition
\begin{equation}
\langle\pi^+(p_2)|V^{\rm em}_{\mu}(0)|
\pi^+(p_1)\rangle =G_\pi (q^2)(p_1+p_2)_{\mu}\,  ,
\end{equation}
we identify the pion formfactor---$G(q^2)=1$. Thus to lowest order in
chiral symmetry the pion has unit charge but is pointlike and structureless.
We shall later see how to insert structure.

\bigskip

{\bf $\pi\pi$ Scattering}\par\bigskip

At two derivative level we can generate additional predictions by
extending our analysis to the case of $\pi\pi$ scattering.  Expanding
${\cal L}_2$ to order $\vec{\phi}^4$ we find
\begin{equation}
{\cal L}_2 : \phi^2={1\over 6v^2} \phi^2\vec{\phi}\cdot \Box\vec{\phi} +
{1\over 2v^2} (\vec{\phi}\cdot\partial_{\mu}\vec{\phi})^2
 + { m^2_\pi \over 24 v^2} \phi^4
\end{equation}
which yields for the pi-pi $T$ matrix
\begin{eqnarray}
T(q_a, q_b,q_c, q_d) & =& {1\over F_{\pi}^2} \left[ \delta^{ab}
\delta^{cd}(s-m^2_{\pi}) +\delta^{ab}\delta^{bd}(t-m_{\pi}^2)
+\delta^{ad}\delta^{bc} (u -m_{\pi}^2)\right]  \nonumber \\
&& -{1\over 3F^2_\pi}(\delta^{ab}\delta^{cd}+\delta^{ac}\delta^{bd}+\delta^{ad}
\delta^{bc})(q^2_a+q_b^2+q_c^2+q_d^2 - 4 m_{\pi}^2)
 \, . \nonumber\\
&&
\end{eqnarray}
Defining more generally
\begin{equation}
 T_{\alpha\beta;  \, \gamma\delta}(s,t,u)=A(s,t,u)
\delta_{\alpha\beta}
\delta_{\gamma\delta}+A(t,s,u)\delta_{\alpha\gamma}
\delta_{\beta\delta}+A(u,t,s)\delta_{\alpha\delta}\delta_{\beta\gamma} \, ,
\end{equation}
we can write the chiral prediction in terms of the more conventional isospin
language by taking appropriate linear combinations\cite{22}
\begin{eqnarray}
T^0(s,t,u)&=& 3A(s,t,u)+A(t,s,u)+A(u,t,s)\, , \nonumber \\
T^1(s,t,u)&=& A(t,s,u)-A(u,t,s)\, , \nonumber \\
T^2(s,t,u)&=& A(t,s,u)+A(u,t,s)\, .
\end{eqnarray}
Partial wave amplitudes, projected out via
\begin{equation}
T_l^I(s)={1\over 64\pi}\int^1_{-1} d(\cos\theta) P_l(\cos \theta)
T^I(s,t,u)  \, ,
\end{equation}
can be used to identify the associated scattering phase shifts as
\begin{equation}
 T^I_l(s) = \left( {s\over s-4m^2_{\pi}}\right)^{1\over 2} e^{i\delta^I_l}
\sin \delta^I_l \,  .
\end{equation}
{}From the lowest order chiral form
\begin{equation}
A(s,t,u)={s-m^2_{\pi}\over F^2_{\pi}}
\end{equation}
we find the predicted values for
the pion scattering lengths and effective ranges
\begin{eqnarray}
a^0_0&=&{7m^2_{\pi}\over 32\pi F^2_{\pi}}\, , \quad
a^2_0=-{m^2_{\pi}\over 16\pi F^2_{\pi}}\, , \quad
a^1_1=-{m^2_{\pi}\over 24\pi F^2_{\pi}}\, ,  \nonumber \\
b^0_0&=&{m^2_{\pi}\over 4\pi F^2_{\pi}}\, , \quad
b^2_0={m^2_{\pi}\over 8\pi F^2_{\pi}}\, ,
\end{eqnarray}
comparison of which with experimentally measured values
is shown in Table 1.

\begin{table}
\begin{center}
\begin{tabular}{l r r r}
\hline \hline
& Experimental & Lowest Order\footnotemark[3]&
First Two Orders\footnotemark[3] \\ \hline
$a^0_0$ & $0.26\pm  0.05 $& 0.16 & 0.20 \\
$b^0_0$ &$ 0.25\pm  0.03$ & 0.18 & 0.26 \\
$a^2_0$ &$-0.028 \pm  0.012$ & -0.045 &-0.041 \\
$b^2_2$ &$-0.082 \pm  0.008$ & -0.089 &-0.070 \\
$a^1_1$ &$0.038 \pm  0.002$ & 0.030 &0.036 \\
$b^1_1$ -- & 0 & 0.043 \\
$a^0_2$ & $(17\pm 3) \times 10^{-4 }$& 0 & $20 \times 10^{-4}$ \\
$a^2_2$ & $(1.3\pm 3) \times 10^{-4 }$& 0 & $3.5 \times 10^{-4}$ \\
\hline\hline
\end{tabular}
\caption{The pion scattering lengths and slopes compared with
predictions of chiral symmetry.}
\end{center}
\end{table}

\par \bigskip

{\bf Difficulties} \par \bigskip
As seen in this Table, these experimental data agree fairly well with the
lowest order theoretical predictions.  However, there exist also
obvious problems, which
show up at higher energy.
\begin{itemize}
\item [i)] Consider first the S-wave I=0 channel
for which
\begin{equation} T^0_0 = {1\over 32\pi F^2_{\pi}}
(2s -m^2_{\pi}) \,  ,\end{equation}
Obviously this form cannot be extended too far in energy since
the unitary condition
\begin{equation} \sqrt{s-4m^2_{\pi} \over s} |T^I_l|^2 \, ,\end{equation}
is violated for $\sqrt{s}>700$ MeV.

\item [ii)] A second indication of problems can be seen in that
unitarity requires the existence of an imaginary component for each
partial wave amplitude
\begin{equation}\mbox{Im} T_l^I=\left({s-4m_\pi^2 \over s} \right)^{1/2}
|T_l^I|^2 \, , \end{equation}
while our simple tree-level analysis yields only real values.
\item [iii)] A third indication of shortcomings is that simple analytic forms
such as result from our analysis cannot possibly reproduce the resonant
behavior seen in $\pi\pi$ scattering such as, for example,
the $\rho$ resonance at 767.0 MeV seen in the P-wave, I=1 channel.
\item [iv)] Our final example of limitations of the simple lowest order chiral
analysis
is provided by the pion electromagnetic form factor.  The unitarity relation
reads in general
\begin{equation}I=S^{\dagger}S=(I-iT^{\dagger})(I+iT) \end{equation}
i.e.
\begin{equation}
i(T-T^{\dagger})=-T^{\dagger} T \, ,
\end{equation}
If we apply this stricture to the $\gamma\pi^+\pi^-$ matrix element
\begin{equation}
i\langle \gamma |T-T^{\dagger}| \pi^+
\pi^- \rangle= -\sum_n \langle
 \gamma|T^{\dagger}|\pi\pi\rangle \langle \pi\pi|
T|\pi^+\pi^-\rangle
\end{equation}
we find
\begin{eqnarray}
&&2 \mbox{Im} G_{\pi}(q^2)(p_1-p_2)_{\mu}  =
\sum {d^3q_1d^3q_2 \over (2\pi)^62 q_1^0
2q_2^0}   \nonumber \\
&  \times & (2 \pi)^4\delta^4(p_1+p_2-q_1-q_2)
(q_1-q_2)_{\mu}\langle\pi^+_{q_1}
\pi^-_{q_2}|T|\pi^+_{p_1}\pi^-_{p_2}
\rangle   \,  ,
\end{eqnarray}
which shows that the unitarity stricture requires an imaginary component
to the pion form factor, in contradiction to our simple result given in
equation

The solution of these problems with unitarity are well known---the
inclusion of loop corrections to these simple tree level
calculations.  Insertion of such loop terms removes the unitarity violations
but comes with a high price---numerous divergences are introduced and this
difficulty
prevented progress in this field for nearly a decade until a paper by
Weinberg suggested the solution.\cite{23}  One can deal with such divergences,
just
as in QED, by introducing counterterms into the Lagrangian in order
to absorb the infinities.
\end{itemize}

\section{Renormalization}

\subsection{Effective Chiral Lagrangian}

We can now apply these lessons to the effective chiral Lagrangian, Eqs. 88,89.
In this case also when loop corrections are made to lowest order
amplitudes in order to enforce unitarity, divergences inevitably arise.
However,
there is an important difference from the familiar case of QED in that the
form of the divergences is {\it different} from their lower order
counterparts.  The reason for this can be seen from a simple example.
Thus consider pi-pi scattering.  In lowest order there exists a
tree level contribution from ${\cal L}_2$ which is ${\cal O} (p^2/F_\pi^2)$
where $p$ represents some generic external energy-momentum.  The fact that
$p$ appears to the second power is due to the feature that its origin is
the two-derivative Lagrangian ${\cal L}_2$.  Now suppose that pi-pi
scattering is examined at one loop order.  Since the scattering amplitude
must still be dimensionless but now the amplitude involves a factor
$1/F_\pi^4$ the numerator must involve {\it four} powers of energy-momentum.
Thus any counterterm which is included in order to absorb this divergence
must be four derivative in character.  Gasser and Leutwyler have studied
this problem and have written the most general form of such an order four
counterterm as\cite{24}

\begin{eqnarray}
{\cal L}_4 &  =&\sum^{10}_{i=1} L_i {\cal O}_i
= L_1\bigg[\tr(D_{\mu}UD^{\mu}U^{\dagger})
\bigg]+L_2\tr (D_{\mu}UD_{\nu}U^{\dagger})\cdot
\tr (D^{\mu}UD^{\nu}U^{\dagger}) \nonumber \\
 &+&L_3\tr (D_{\mu}U D^{\mu}U^{\dagger}
D_{\nu}U D^{\nu}U^{\dagger})
+L_4 \tr  (D_{\mu}U D^{\mu}U^{\dagger})
\tr (\chi{U^{\dagger}}+U{\chi}^{\dagger}
) \nonumber \\
&+&L_5\tr \left(D_{\mu}U D^{\mu}U^{\dagger}
\left(\chi U^{\dagger}+U \chi^{\dagger}\right)
\right)+L_6\bigg[ \tr \left(\chi U^{\dagger}+
U \chi^{\dagger}\right)\bigg]^2 \nonumber \\
&+&L_7\bigg[ \tr \left(\chi^{\dagger}U-
U\chi^{\dagger}\right)\bigg]^2 +L_8 \tr
\left(\chi U^{\dagger}\chi U^{\dagger}
+U \chi^{\dagger}
U\chi^{\dagger}\right)\nonumber \\
&+&iL_9 \tr \left(F^L_{\mu\nu}D^{\mu}U D^{\nu}
U^{\dagger}+F^R_{\mu\nu}D^{\mu} U^{\dagger}
D^{\nu} U \right) +L_{10} \tr\left(F^L_{\mu\nu}
U F^{R\mu\nu}U^{\dagger}\right) \nonumber\\
\end{eqnarray}
where the constants $\alpha_i, i=1,2,\ldots 10$ are arbitrary and
$F^L_{\mu\nu}, F^R_{\mu\nu}$ are external field strength tensors defined via
\begin{eqnarray}
F^{L,R}_{\mu\nu}=\partial_\mu F^{L,R}_\nu-\partial_\nu
F^{L,R}_\mu-i[F^{L,R}_\mu ,F^{L,R}_\nu],\qquad F^{L,R}_\mu =V_\mu\pm A_\mu
\end{eqnarray}
Now just as in the case of QED the bare parameters $L_i$ which appear
in this Lagrangian are not physical.  Rather the ``physical" (renormalized)
values of these parameters are obtained by appending to these bare values
the divergent one-loop contributions having the appropriate form
\begin{equation} L^r_i = L_i -{\gamma_i\over 32\pi^2}
\left[{-2\over \epsilon} -\ln (4\pi)+\gamma -1\right]\end{equation}
By comparing with experiment, Gasser and Leutwyler have determined experimental
values for each of these ten parameters.  While ten sounds like a rather
large number, we shall see below that this picture is actually {\t predictive}.
Typical values for the parameters are shown in Table 2.
\begin{table}
\begin{center}
\begin{tabular}{l l c}\hline\hline
Coefficient & Value & Origin \\
\hline
$L_1^r$ & $0.65\pm 0.28$ & $\pi\pi$ scattering \\
$L_2^r$ & $1.89\pm 0.26$ & and\\
$L_3^r$ & $-3.06\pm 0.92$ & $K_{\ell 4}$ decay \\
$L_5^r$ & $2.3\pm 0.2$ & $F_K/F_\pi$\\
$L_9^r$ & $7.1\pm 0.3$ & $\pi$ charge radius \\
$L_{10}^r$ & $-5.6\pm 0.3$ & $\pi\rightarrow e\nu\gamma$\\
\hline\hline
\end{tabular}
\caption{Gasser-Leutwyler counterterms and the means by which
they are determined.}
\end{center}
\end{table}

The question which one should ask at this point is why stop at order four?
Clearly if two loop corrections from ${\cal L}_2$ or one-loop
corrections from ${\cal L}_4$ are calculated, divergences will arise which
are of six derivative character.  Why not include these?  The answer is that
the chiral procedure represents an expansion in energy-momentum.  Corrections
to the tree level predictions from one loop corrections from ${\cal L}_2$
or tree level contributions from ${\cal L}_4$ are ${\cal
O}(E^2/\Lambda_\chi^2)$
where $\Lambda_\chi\sim 4\pi F_\pi\sim 1$ GeV is the chiral scale.\cite{25}
Thus chiral
perturbation theory is a {\it low energy} procedure.  It is only to the extent
that the energy is small compared to the chiral scale that it makes sense to
truncate the expansion at the four-derivative level.  Realistically this
means that we deal with processes involving $E<500$ MeV, and, as we shall
describe below, for such reactions the procedure is found to work very well.

Now let's give an example of a chiral perturbation theory calculation
in order to see how it is performed and in order to see how the experimental
counterterm values are actually determined.  Consider the pion
electromagnetic form factor, which by Lorentz- and gauge-invariance has the
structure
\begin{equation} \langle \pi^+(p_2)|J^{\mu}_{\mbox{em}}|\pi^+(p_1)
\rangle=
G_{\pi}(q^2)(p_1+p_2)^{\mu} \end{equation}
We begin by identifying the electromagnetic current as
\begin{eqnarray}
J^{\mu}_{\mbox{em}}& =& -{\partial{\cal L}\over \partial
(eA_{\mu})}=(\varphi \times \partial^{\mu}\varphi)_3\bigg[
1-{1\over 3F^2}\varphi\cdot \varphi+{\cal O}(\varphi^4)\bigg]
\nonumber \\
&+&(\varphi\times\partial^{\mu}\varphi)_3
\bigg[16L_4
+8L_5\bigg]{m^2_{\pi}\over F^2}+
{4L_9\over F^2}\partial^{\nu}(\partial^{\mu}
\varphi\times\partial_{\nu}\varphi)_3+
\cdots
\end{eqnarray}
where we have expanded to {\it fourth} order in the pseudoscalar fields.
Defining

\begin{eqnarray}
\delta_{jk}I(m^2)&=&i\Delta_{Fjk}(0)=\langle 0|\varphi_j(x)
\varphi_k(x)|0\rangle , \nonumber \\
I(m^2)&=&\mu^{4-d}\int{d^d k\over (2\pi)^d}{i\over
 k^2-m^2}={\mu^{4-d}\over (4\pi)^{d/2}} \Gamma
\left(1-{d\over 2}\right)(m^2)^{{d\over 2}-1} , \nonumber \\
\delta_{jk}I_{\mu\nu}(m^2) &=& -\partial_{\mu}\partial_{\nu}
i\Delta_{Fjk}(0)=\langle 0|\partial_{\mu}\varphi_j(x)\partial_{\nu}
\varphi_k(x)|0\rangle , \nonumber \\
I_{\mu\nu}(m^2) & =& \mu^{4-d} \int{d^d k\over (2\pi)^d }
k_{\mu}k_{\nu} {i\over k^2 -m^2}=g_{\mu\nu}{m^2\over d}
I(m^2)
\end{eqnarray}
we calculate the one loop correction shown in Figure 2a to be

\begin{equation} J^{\mu}_{\mbox{em}}|_{(2a)}= -{5\over
 3F^2_{\pi}}
(\varphi\times\partial^{\mu}\phi)_3I
(m^2_{\pi}) \end{equation}

\begin{figure}
\vspace{5cm}
\caption{Loop corrections to the pion form factor.}
\end{figure}

We also need the one loop correction shown in Figure 2b.  For this
piece we require the form of the pi-pi scattering amplitude which
arise from ${\cal L}_2$

\begin{equation}
\langle\pi^+(k_1)\pi^-(k_2)|\pi^+(p_1)\pi^-(p_2)\rangle
= {i\over 3F_0^2}\left(2m^2_0+p^2_1+p^2_2+
k^2_1+k^2_2-3(p_1-k_1)^2\right)
\end{equation}
and we use the results of dimensional integration
\begin{eqnarray}
& &\int {d^d p\over (2\pi)^d}
{1\over [(p-q)^2-
m^2_1+i\varepsilon]^{n_1}[p^2-m^2_2+i
\varepsilon]^{n_2}} \nonumber \\
&=& (-1)^{n_1+n_2}{i\over (4\pi)^{d\over 2}}{\Gamma(n_1+
n_2-d/2)\over \Gamma(n_1)\Gamma(n_2)}
\int^1_0 dx
{x^{n_1-1}(1-x)^{n_2-1}\over{\cal D}^{n_1+n_2-d/2}}\nonumber
\end{eqnarray}

\begin{eqnarray}
& &\int {d^d p\over (2\pi)^d}
{p^{\mu}\over [(p-q)^2-
m^2_1+i\varepsilon]^{n_1}[p^2 - m^2_2 +i
\varepsilon]^{n_2}} \nonumber \\
&=& (-1)^{n_1+n_2}q^{\mu}{i\over (4\pi)^{d\over
 2}}{\Gamma  (n_1+
n_2-d/2)\over \Gamma(n_1)\Gamma(n_2)} \int^1_0 dx
{x^{n_1}(1-x)^{n_2-1}\over {\cal D}^{n_1+n_2-d/2}}
 \nonumber
\end{eqnarray}

\begin{eqnarray}
& &\int {d^d p\over (2\pi)^d}{p^{\mu}p^{\nu}
\over [(p-q)^2- m^2_1+i\varepsilon]^{n_1}[p^2-m^2_2+i
\varepsilon]^{n_2}} \nonumber \\
&=& {i\over (4\pi)^{d\over 2}}{(-1)^{n_1+n_2}\over \Gamma(n_1)\Gamma(n_2)}
\bigg[q^{\mu}q^{\nu}\Gamma
(n_1+n_2-d/2) \int^1_0 dx
{x^{n_1+1}(1-x)^{n_2-1}\over {\cal D}^{n_1+n_2-d/2}}
 \nonumber \\
& & -{g^{\mu\nu}\over 2} \Gamma(n_1+n_2-1-d/2)
\int^1_0 dx{x^{n_1-1}(1-x)^{n_2-1}\over {\cal D}^{n_1+
n_2-1-d/2}}\bigg]
\end{eqnarray}
where
\begin{equation} {\cal D} \equiv m^2_1 x +m^2_2(1-x)-q^2
 x(1-x)-i\varepsilon \end{equation}
In the limit
$d\rightarrow 4$ we find that
\begin{equation} \Gamma\left(2-{d\over2}\right)={1\over \varepsilon}
-\gamma + {\cal O}(\varepsilon) \end{equation}
The integration in Figure 2b may now be done, yielding
\begin{eqnarray}
\langle J^{\mu}_{\mbox{em}}\rangle _{(2b)} &=& {1\over (4\pi
F_{\pi})^2} (p_1+p_2)^{\mu} \int^1_0 dx(m^2_{\pi} -q^2
 x(1-x)) \nonumber \\
&\times& \bigg[\left(-{2\over \epsilon}+\gamma-1-\ln 4\pi\right)
+\ln {m^2_{\pi}-q^2 x(1-x)\over \mu^2 }\bigg]
\end{eqnarray}
Performing the x-integration we find, finally
\begin{eqnarray}
\langle J^{\mu}_{\mbox{em}}\rangle _{(2b)} &=& {1\over (4\pi
F_{\pi})^2} (p_1+p_2)^{\mu} \bigg\{\left(m^2_{\pi}-
{1\over 6} q^2 \right) \bigg[ -{2\over \epsilon} +\gamma -1
-\ln 4\pi +\ln {m^2_{\pi}\over \mu^2}\bigg] \nonumber  \\
&& + {1\over 6} (q^2-4m^2_{\pi})H\left({q^2\over
m^2_{\pi}}\right)-{1\over 18} q^2\bigg\}
\end{eqnarray}
Here the function H(a) is given by
\begin{eqnarray}
H(a) & \equiv & \int_0^1 dx \ln (1-a x(1-x)) \nonumber \\
& = & \Bigg\{\begin{array}{l l}
2-2\sqrt{{4\over a}-1} \cot^{-1}\sqrt{{4\over a}-1}& (0<a<4) \\
2+\sqrt{1-{4\over a}}\bigg[\ln {\sqrt{1-{4\over a}}-1 \over
\sqrt{1-{4\over a}} +1} + i\pi \theta (a-4) \bigg] &
\mbox{(otherwise) }
\end{array}%\right.
\end{eqnarray}
and contains the imaginary component required by unitarity.

We are not done yet, however, since we must also include mass and
wavefunction effects.  In order to do so, we expand ${\cal L}_2$
to fourth order in $\varphi(x)$, and ${\cal L}_4$ to second order:

\begin{eqnarray}
{\cal L}_2&=&{1\over 2} [\partial^{\mu} \varphi\cdot
 \partial_{\mu} \varphi -m^2_0\varphi\cdot\varphi]+
{m^2_0\over 24 F^2_0}(\varphi\cdot\varphi)^2 \nonumber \\
&& +{1\over 6F^2_0}[(\varphi\cdot\partial^{\mu}
\varphi)(\varphi\cdot\partial_{\mu}\varphi)-(\varphi
\cdot\varphi)(\partial^{\mu}\varphi\cdot\partial_{\mu}
\varphi)]+{\cal O}(\varphi^6), \nonumber \\
{\cal L}_4&=&{m^2_0 \over F^2_0}[16L_4
+8L_5]{1\over 2}\partial_{\mu}
\varphi\cdot\partial^{\mu}\varphi \nonumber \\
&& -{m^2_0\over F^2_0}[32L_6+16
L_8]{1\over 2} m^2_0\varphi\cdot\varphi+
{\cal O}(\varphi^4) .
\end{eqnarray}
Performing the loop integrations on the $\phi^4(x)$ component of the above
yields
\begin{eqnarray}
{\cal L}_{\mbox{eff}}& =& {1\over 2} \partial^{\mu} \varphi
\partial_{\mu} \varphi -{1\over 2} m^2_0 \varphi \cdot
 \varphi +{5m^2_{\pi} \over 12F^2_{\pi}}
 I(m^2_{\pi})\varphi\cdot \varphi \nonumber \\
&&+{1\over 6F^2_{\pi}}(\delta_{ik}\delta_{jl}-\delta_{ij}
\delta_{kl}) I(m^2_{\pi})(\delta_{ij}\partial^{\mu}\varphi_k
\partial_{\mu} \varphi_l+\delta_{kl}m^2_{\pi}\varphi_i\varphi_j)
\nonumber  \\
&& +{1\over 2} \partial_{\mu}\varphi\cdot \partial^{\mu}
\varphi{m^2_{\pi}\over F^2_{\pi}}[16L_4+8
L_5)]-{1\over 2}m^2_{\pi}\varphi\cdot\varphi
{m^2_{\pi}\over F^2_{\pi}}[32L_6+16
L_8\bigg] \nonumber \\
&=&{1\over 2}\partial^{\mu}\varphi\cdot\partial_{\mu}
\varphi \bigg[1+(16L_4+8L_5)
{m^2_{\pi}\over F^2_{\pi}} -{2\over 3F^2_{\pi}}I(m^2_{\pi})
\bigg] \nonumber \\
&& -{1\over 2} m^2_0\varphi\cdot\varphi \bigg[
 1+(32L_6+16L_8){m^2_{\pi}\over
F^2_{\pi}} -{1\over 6F^2_{\pi}}I(m^2_{\pi})\bigg]
\end{eqnarray}
from which we can now read off the wavefunction renormalization
term $Z_\pi$.

When this is done we find

\begin{eqnarray}
Z_{\pi}G_{\pi}^{\mbox{(tree)}}(q^2)&=&
\bigg[1-{8m^2_{\pi}\over
 F^2_{\pi}}(2L_4 +L_5 \nonumber \\
& & +{m^2_{\pi}\over 24\pi^2 F^2_{\pi}}
\bigg\{-{2\over \epsilon}+\gamma -1-\ln
4\pi +\ln {m^2_{\pi}\over \mu^2}\bigg\}\bigg]
 \nonumber \\
& & \times\bigg[1+{8m^2_{\pi}\over F^2_{\pi}}
(2L_4+L_5)+2 q^2{L_9
\over F^2_{\pi}} \bigg] \nonumber  \\
&=& \bigg[ 1+{m^2_{\pi}\over 24\pi^2
 F^2_{\pi}} \bigg(-{2
\over \epsilon} +\gamma -1 -\ln 4\pi +\ln {m^2_{\pi}\over
 \pi^2}\bigg)+{2L_9\over F^2_{\pi}}q^2 \bigg]\nonumber\\
\end{eqnarray}
while from the loop diagrams given earlier
\begin{eqnarray}
G_{\pi}(q^2)\Bigg|_{(2a)}&=& -{5m^2_{\pi}\over 48\pi^2
F^2_{\pi}}\Bigg\{-{2\over \epsilon}+\gamma-1-\ln 4\pi
 +\ln{m^2_{\pi}\over \mu^2}\Bigg\} \nonumber \\
G_{\pi}(q^2)\Bigg|_{(2b)}&=& {1\over 16\pi^2 F^2_{\pi}}
\Bigg\{\Bigg(m^2_{\pi}-{1\over 6}q^2\Bigg)
\Bigg[ -{2\over \epsilon}
+\gamma -1 -\ln 4\pi +\ln {m^2_{\pi}\over \mu^2}\Bigg]
\nonumber \\
&& \left.
+{1\over 6}(q^2-4m^2_{\pi})H\left({q^2\over m^2_{\pi}}
\right)-{1\over 18}q^2\right\}
\end{eqnarray}
Adding everything together we have the final result, which when
written in terms of the renormalized value $L_9^{(r)}$ is {\it finite}!
\begin{eqnarray}
G_{\pi}(q^2) &=& 1+{2L_9^{(r)}\over
 F^2_{\pi}}q^2
+{1\over 96\pi^2F^2_{\pi}}
\left[(q^2-4m^2_{\pi})H\left({q^2\over
 m^2_{\pi}}\right)-q^2\ln {m^2_{\pi}\over
 \mu^2}-{q^2\over 3}\right] \nonumber\\
&&
\end{eqnarray}
If we expand to lowest order in $q^2$ we find
\begin{eqnarray}
G_{\pi}(q^2) &=& 1+q^2\left[
{2L^{(r)}_9\over F^2_{\pi}} -{1\over 96
\pi^2F^2_{\pi}}\left(\ln {m^2_{\pi}\over \mu^2}+
1\right)\right]+\cdots
\end{eqnarray}
which can be compared with the phenomenological description in terms of
the pion charge radius
\begin{equation} G_{\pi}(q^2)=1+{1\over 6}\langle
 r^2_{\pi}\rangle q^2 +\cdots \end{equation}
By equating these two expressions and using the experimental value
of the pion charge radius---$\langle r_\pi^2\rangle_{\rm exp}=
(0.44\pm 0.01) \mbox{fm}^2$\cite{26}---we determine the value of the
counterterm $L_9^{(r)}$ shown in Table 2.

Although due to lack of space, we have in these lectures limited our
attention to only a very limited number of reactions, chiral perturbative
techniques have been applied to many other processes involving Goldstone
interaction, some examples of which are indicated in Table 2.
\subsection{Anomalous Symmetry Breaking}

Thus far we have given examples of {\it explicit} and of {\it spontaneous}
symmetry breaking within QCD.  Remarkably QCD
also involves {\it anomalous} symmetry breaking, which can be characterized
in terms of an effective Lagrangian which has {\it no} free parameters.
The form of this interaction has been given by Witten as\cite{27}
\begin{equation} {\cal L}_{\mbox{A}}= -{N_c\over 48\pi^2}\varepsilon^{\mu
\nu\alpha\beta}[eA_{\mu}\tr (QL_{\nu}L_{\alpha}
L_{\beta}-QR_{\nu}R_{\alpha}R_{\beta})+ie^2
F_{\mu\nu}A_{\alpha}T_{\beta} ] \end{equation}
where
\begin{eqnarray}
L_{\mu} &\equiv& \partial_{\mu}U U^{\dagger},
\qquad
R_{\mu}\equiv \partial_{\mu}U^{\dagger}U,
\nonumber \\
T_{\beta}&=& \tr \left( Q^2L_{\beta}-Q^2R_{\beta}
+{1\over 2}Q U Q U^{\dagger}L_{\beta}-{1\over 2}
Q U^{\dagger}Q U R_{\beta}\right)
\end{eqnarray}
and $N_c$ is the number of colors.  The best known manifestation of
the anomaly is its prediction for the process
$\pi^0\rightarrow\gamma\gamma$.  The component of ${\cal L}_{\rm A}$
which is responsible for this process can be identified as

\begin{equation} {\cal L}_A = {e^2N_c \over 48\pi^2F_{\pi}}
3\tr (Q^2\tau_3)\varepsilon^{\mu\nu\alpha\beta}
F_{\mu\nu}
A_{\alpha}\partial_{\beta}\pi^0=
{\alpha N_c \over
24\pi F_{\pi}}\varepsilon^{\mu\nu\alpha\beta}
F_{\mu\nu}F_{\alpha\beta}\pi^0\end{equation}
Defining the decay amplitude as
\begin{equation} {\cal M}_{\pi^0\rightarrow\gamma\gamma}
=-iA_{\gamma\gamma}\varepsilon^{\mu\nu
\alpha\beta}\varepsilon^{\ast}_{\mu}k_{\nu}
\varepsilon'^{\ast}_{\alpha}k'_{\beta} \end{equation}
we find the decay rate
\begin{equation} \Gamma_{\pi^0\rightarrow \gamma\gamma}
= {m^3_{\pi}\over 64 \pi} |A_{\gamma
\gamma}|^2 \end{equation}
The decay amplitude predicted by the anomaly is found to be
\begin{equation} A_{\gamma\gamma}={\alpha N_c\over 3\pi
 F_{\pi}} \stackrel{N_c=3}{\Longrightarrow} 0.025
 \, \mbox{GeV}^{-1} \end{equation}
which is in excellent agreement with the value determined by
experiment\cite{28}
\begin{equation} A_{\gamma\gamma}= 0.0025 \pm 0.001\,
\mbox{GeV}^{-1} \end{equation}
and gives eloquent proof that the number of colors is precisely three.
Although this is the best known example, there are numerous additional
manifestations of the anomaly, {\it e.g.} $\gamma \longrightarrow  3\pi,
K  \longrightarrow
\pi\pi e\nu_e,
\eta \longrightarrow  \pi\pi\gamma,$ {\it etc.}

\subsection{Comparison with Experiment}
Of course, having gone to such effort to set up this formalism, the
real question is ``Does it make successful predictions?"  We do not have the
space here to give
a detailed answer to this question, so a simple example will have to suffice.
A particularly interesting indication of the predictive power of chiral
perturbation theory is found in the semileptonic weak process
$K \rightarrow \pi\pi l \nu_l $
for which one defines the matrix element\cite{29}
\begin{eqnarray}
 \langle \pi^+(p_+)\pi^-(p_-) |\bar{s}
 \gamma_{\mu} \gamma_5 u| K^+(k)\rangle  &=&
-{i \over \sqrt{2}F_{\pi}} \nonumber \\
\times \left[(p_++p_-)_{\mu}f_1+(p_+-p_-)_{\mu}\right.
f_2 &+&\left.(k-p_+-p_-)_{\mu} f_3 \right] \nonumber \\
\langle\pi^+(p_+)\pi^-(p_-) |\bar{s}
 \gamma_{\mu}  u| K^+(k)\rangle &=&
{2 g\over \pi F^3_{\pi}} \varepsilon_{\mu
\nu \alpha \beta} k^{\nu} p^{\alpha}_+
p^{\beta}_-
\end{eqnarray}
Parameterizing the form factors as
\begin{equation} f_i(k^2)= f_i(0) \left[1+\lambda_i{k^2\over
m^2_{\pi}}\right] \quad {\rm with} \quad
k^2={1\over 4} \left((p_++p_-)^2-4m^2_{\pi}
\right) \end{equation}
one finds predicted values which are in excellent agreement with
those determined experimentally as shown in Table 3.
Notice also that higher order corrections are essential here.

\begin{table}
\begin{center}
\begin{tabular}{l l l l}\hline\hline
 &Data & Lowest Order & Order (E$^4$) \\
\hline
$f_1(0)$& $1.47\pm 0.04$& 1.00& 1.45 \\
$f_2(0)$& $1.25\pm 0.07$& 1.00&1.24 \\
$\lambda_1$& $0.08\pm 0.02$& 0.00& 0.08 \\
$\lambda_2$& $0.08\pm 0.02$& 0.00& 0.06 \\
$g(0)$&$0.96\pm 0.24$& 0.00& 1.00\\
\hline\hline
\end{tabular}
\caption{Experimental values of $K_{\ell 4}$ parameters and their chiral
predictions.}
\end{center}
\end{table}
A second example of predictive power can be seen by examining the connection
between radiative pion decay and pion Compton scattering.  For the former
we can define\cite{30}
\begin{eqnarray}
{\cal M}_{\pi^+\rightarrow e^+\nu_e\gamma}
&=&-{eG_F\over \sqrt{2}} \cos\theta_1 M_{\mu
\nu}(p,q)\varepsilon^{\mu\ast}(q)\bar{u}(p_{\nu})
\gamma^{\nu}(1+\gamma_5)v(p_e) \nonumber \\
{\cal M}_{\pi^+\rightarrow e^+\nu_e e^+ e^-}
&=&-{e^2G_F\over \sqrt{2}} \cos\theta_1 M_{\mu
\nu}(p,q){1\over q^2} \nonumber \\
&& \times \bar{u}(p_2)
\gamma^{\mu} v(p_1)\bar{u}(p_{\nu}) \gamma^{\nu}
 (1+\gamma_5)v(p_e)
\end{eqnarray}
where the hadronic component of $M_{\mu\nu}$ has the structure
\begin{eqnarray}
M_{\mu\nu}(p,q)&=&\int d^4xe^{iq\cdot x}<0|T(J^{\rm em}_\mu (x)
J_\nu^{1-i2}(0)|\pi(\vec{p})>
= \mbox{Born terms}\nonumber\\
&-&h_A((p-q)_\mu q_\nu -g_{\mu\nu}q\cdot (p-q))
-r_A(q_\mu q_\nu -g_{\mu\nu}q^2)\nonumber\\
&+&ih_V\epsilon_{\mu\nu\alpha\beta}q^\alpha
p^\beta
\end{eqnarray}
where $h_A,r_A,h_V$ are unknown structure functions.  (Note that $r_A$
can only be measured via the rare Dalitz decay $\pi^+\rightarrow e^+
\nu_e e^+e^-$.)  Likewise we can define the amplitude for Compton
scattering as
\begin{eqnarray}
-iT_{\mu\nu}(p,p',q)&=&-i\int d^4x e^{iq_1\cdot x} <\pi^+(\vec{p}')|
T(J_\mu^{\rm em}(x)J_\nu^{\rm em}(0)|\pi^+(\vec{p})>\nonumber\\
&=&\mbox{Born terms}+\sigma(q_{2\mu}q_{1\nu}-g_{\mu\nu}q_1\cdot q_2)+\cdots
\end{eqnarray}
Chiral symmetry makes {\it four} predictions among these parameters,
and three of the four are found to be in good agreement
with experiment.  The possible exception involves a relation between the
charged
pion polarizability and the axial structure constant $h_A$ measured in
radiative pion decay.  In this case there exist two conflicting
experimental results, one of which agrees and one of which does not agree
with the theoretical prediction.  It is important to resolve this potential
discrepancy, since these chiral predictions are firm ones.  There is no
way (other than introducing perversely large higher order effects) to
bring things into agreement were some large violation of a chiral prediction
to be verified, since the only ingredient which goes into such predictions
is the chiral symmetry of QCD itself!

\section{Baryon Chiral Perturbation Theory}

Our discussion of chiral methods given above was limited to the study of
the interactions of the pseudoscalar mesons with leptons and with each
other.  In the real world, of course, interactions with baryons also
take place and it is an important problem to develop a useful predictive
scheme based on chiral invariance for such processes.  Again much work
has been done in this regard, but there remain important problems.  Writing
down the lowest order chiral Lagrangian at the SU(2) level is
straightforward---
\begin{equation}
{\cal L}_{\pi N}=\bar{N}(i\rlap /{D}-m_N+{g_A\over 2}\rlap /{u}\gamma_5)N
\end{equation}
where $g_A$ is the usual nucleon axial coupling in the chiral limit, the
covariant derivative $D_\mu=\partial_\mu+\Gamma_\mu$ is given by
\begin{equation}
\Gamma_\mu={1\over 2}[u^\dagger,\partial_\mu u]-{i\over 2}u^\dagger
(V_\mu+A_\mu)u-{i\over 2}u(V_\mu-A_\mu)u^\dagger
\end{equation}
and $u_\mu$ represents the axial structure
\begin{equation}
u_\mu=iu^\dagger\nabla_\mu Uu^\dagger
\end{equation}
Expanding to lowest order we find
\begin{eqnarray}
{\cal L}_{\pi N}&=&\bar{N}(i\rlap /{\partial}-m_N)N+{g_A\over F_\pi}
\bar{N}\gamma^\mu\gamma_5{1\over 2}\vec{\tau}N\cdot\vec{\pi}\nonumber\\
&-&{1\over 4F_\pi^2}\bar{N}\gamma^\mu\vec{\tau}N\cdot\vec{\pi}\times
\partial\vec{\pi}+\ldots
\end{eqnarray}
which yields the Goldberger-Treiman relation,
connecting strong and axial couplings of the nucleon system\cite{50}
\begin{equation}
F_\pi g_{\pi NN}=m_N g_A
\end{equation}
Using the present best values for these quantities, we find
\begin{equation}
92.4 \mbox{MeV}\times 13.0 =1201 \mbox{MeV}\quad\mbox{vs.}\quad 1183 \mbox{MeV}
= 939\mbox{MeV}\times 1.26
\end{equation}
The agreement to better than two percent strongly confirms the validity
of chiral symmetry in the nucleon sector.

\subsection{Heavy Baryon Methods}

Extension to SU(3) gives additional successful predictions---the linear
Gell-Mann-Okubo relation as well as the generalized Goldbeger-Treiman
relation.  However, difficulties arise when one attempts to include
higher order corrections to this formalizm.  The difference from the
Goldstone case is that there now exist {\it two} dimensionful
parameters---$m_N$ and $F_\pi$---in the problem rather than
{\it one}---$F_\pi$.  Thus loop effects can be of order $(m_N/4\pi F_\pi)^2
\sim 1$ and we no longer have a reliable perturbative scheme.  A
consistent power counting mechanism can be constructed provided that we
eliminate the nucleon mass from the Lagrangian.  This is done by
considering the nucleon to be very heavy.  Then we can write its
four-momentum as\cite{51}
\begin{equation}
p_\mu=Mv_\mu+k_\mu
\end{equation}
where $v_\mu$ is the four-velocity and satisfies $v^2=1$, while $k_\mu$
is a small off-shell momentum, with $v\cdot k<< M$.  One can then construct
eigenstates of the projection operators $P_\pm = {1\over 2}(1\pm
\rlap /{v})$, which in the rest frame project out upper, lower
components of the Dirac wavefunction, so that\cite{52}
\begin{equation}
\psi=e^{-iMv\cdot x}(H_v+h_v)
\end{equation}
where
\begin{equation}
H_v=P_+\psi,\qquad h_v=P_-\psi
\end{equation}
The effective Lagrangian can then be written in terms of $N,h$ as
\begin{equation}
{\cal L}_{\pi N}=\bar{H}_v{\cal A}H_v+\bar{h}_v{\cal B}H_v+
\bar{H}_v\gamma_0{\cal B}^\dagger\gamma_0h_v-\bar{h}_v{\cal C}h_v
\end{equation}
where the operators ${\cal A}, {\cal B},{\cal C}$ have the low energy
expansions
\begin{eqnarray}
{\cal A}&=&iv\cdot D+g_A u\cdot S +\ldots\nonumber\\
{\cal B}&=&i\rlap /{D}^\perp-{1\over 2}g_A v\cdot u\gamma_5+\ldots\nonumber\\
{\cal C}&=&2M+iv\cdot D+g_A u\cdot S+\ldots
\end{eqnarray}
Here $D_\mu^\perp=(g_{\mu\nu}-v_\mu v_\nu)D^\nu$ is the transverse component
of the covariant derivative and $S_\mu={i\over 2}\gamma_5
\sigma_{\mu\nu}v^\nu$ is the Pauli-Lubanski spin vector and satisfies
the relations
\begin{equation}
S\cdot v=0,\quad S^2=-{3\over 4},\quad\{S_\mu,S_\nu\}={1\over 2}(v_\mu v_\nu-
g_{\mu\nu}),\quad [S_\mu,S_\nu]=i\epsilon_{\mu\nu\alpha\beta}v^\alpha S^\beta
\end{equation}
We see that the two components H,h are coupled in this expression for the
effective action.  However, this may be undone by the field transformation
\begin{equation}
h'=h-{\cal C}^{-1}{\cal B}H
\end{equation}
in which case the Langrangian becomes
\begin{equation}
{\cal L}_{\pi N}=\bar{H}_v({\cal A}+(\gamma_0{\cal B}^\dagger
\gamma_0){\cal C}^{-1}{\cal B})H_v-\bar{h}'_v{\cal C}h'_v
\end{equation}
The piece of the Lagrangian involving $H$ no longer contains the mass as a
parameter and is the effective Lagrangian that we desire.  The remaining
piece involving $h'_v$ can be thrown away, as it does not couple to the
$H_v$ physics.  (In path integral language we simply integrate out this
component yielding an uninteresting overall constant.)
Of course, when loops are calculated a set of
counterterms will be required and these are given at leading (two-derivative)
order by
\begin{eqnarray}
{\cal A}^{(2)}&=&{M\over F_\pi^2}(c_1\mbox{Tr}\chi_+
+c_2(v\cdot u)^2+c_3u\cdot u+c_4[S^\mu,s^\nu]u_\mu u_\nu\nonumber\\
&+&c_5(\chi_+-\mbox{Tr}\chi_+)
-{i\over
4M}[S^\mu,S^\nu]((1+c_6)F^+_{\mu\nu}+c_7\mbox{Tr}f^+_{\mu\nu}))\nonumber\\
{\cal B}^{(2)}&=&{M\over F_\pi^2}((-{c_2\over 4}i[u^\mu,u^\nu]+c_6f_+^{\mu\nu}
+c_7Trf_+^{\mu\nu})\sigma_{\mu\nu}\nonumber\\
&-&{c_4\over 2}v_\mu\gamma_\nu Tru^\mu u^\nu)\nonumber\\
{\cal C}^{(2)}&=&-{M\over F_\pi^2}(c_1Tr\chi_++(-{c_2\over 4}i[u^\mu,u^\nu]
+c_6f_+^{\mu\nu}+c_7trF_+^{\mu\nu})\sigma_{\mu\nu}\nonumber\\
&-&{c_3\over 4}Tr u^\mu u_\nu
-({c_4\over 2}+Mc_5)v_\mu v_\nu Tru^\mu u^\nu)
\end{eqnarray}
Expanding ${\cal C}^{-1}$ and the other terms in terms of
a power series in $1/M$ then leads to an
effective heavy nucleon Lagrangian of the form (to ${\cal O}(q^3)$)
\begin{eqnarray}
{\cal L}_{\pi N}&=&\bar{H}_v\{{\cal A}^{(1)}+{\cal A}^{(2)}+{\cal A}^{(3)}
+(\gamma_0{\cal B}^{(1)\dagger}\gamma_0){1\over 2M}{\cal B}^{(1)}\nonumber\\
&+&{(\gamma_0{\cal B}^{(1)\dagger}\gamma_0){\cal B}^{(2)}+(\gamma_0{\cal B}
^{(2)\dagger}\gamma_0){\cal B}^{(1)}\over 2M}\nonumber\\
&-&(\gamma_0{\cal B}^{(1)\dagger}\gamma_0){i(v\cdot D)+g_A(u\cdot S)\over
(2M)^2}{\cal B}^{(1)}\}H_v+{\cal O}(q^4)
\end{eqnarray}
A set of Feynman rules can now be written down and a consistent power counting
scheme developed, as shown by Meissner and his collaborators.\cite{53}

\subsection{Applications}
As our first example consider the nucleon-photon interaction.  To lowest (one
derivative) order we have from ${\cal A}^{(1)}$
\begin{equation}
{\cal L}_{\gamma NN}^{(1)}=ie\bar{N}{1\over2}(1+\tau_3)\epsilon\cdot vN
\end{equation}
while at two-derivative level we find
\begin{eqnarray}
 {\cal L}_{\gamma NN}^{(2)}=\bar{N}\left\{{e\over
4M}(1+\tau_3)\epsilon\cdot(p_1+p_2)
+{ie\over 2M}[S\cdot \epsilon,S\cdot k](1+\kappa_S+\tau_3(1+\kappa_V)\right\}N
\nonumber\\
\quad
\end{eqnarray}
whre we have made the identifications
$c_6=\kappa_V,\quad c_7={1\over 2}(\kappa_S-\kappa_V)$.  We can now reproduce
the low energy theorems for Compton scattering.  Consider the case of the
proton.
At the two derivative level,
we have the tree level prediction from
\begin{equation}
(\gamma_0{\cal B}^{(1)\dagger}\gamma_0){1\over 2M}
{\cal B}^{(1)}|_{\gamma pp}={e^2\over 2M}\vec{A}_\perp^2
\end{equation}
which yields the familiar Thomson amplitude
\begin{equation}
\mbox{Amp}_{\gamma pp}=-{e^2\over M}\hat{\epsilon}'\cdot\hat{\epsilon}
\end{equation}
On the other hand at order $q^3$ we find a contribution from the pole diagrams
shown in Figure 3 with two-derivative terms at each vertex.  This yields
\begin{eqnarray}
\mbox{Amp}_{\gamma pp}&=&({e\over M}^2){1\over \omega}
\bar{p}[(\hat{\epsilon}'
\cdot\vec{k}\vec{S}\cdot\hat{\epsilon}\times\vec{k}
-\hat{\epsilon}\cdot\vec{k}'\vec{S}\cdot
\hat{\epsilon}'\times\vec{k}')(1+\kappa_p)\nonumber\\
&+&i\vec{S}\cdot(\hat{\epsilon}\times\vec{k})\times
(\hat{\epsilon}'\times\vec{k}')(1+\kappa_p)^2]
\end{eqnarray}
The full result must also include contact terms at order $q^3$ from the last
piece of Eq. 171
\begin{equation}
-eP_+\rlap /{A}^\perp {iv\cdot D\over (2M)^2}e\rlap /{A}^\perp P_+=
-{e^2\over 2M^2}\vec{S}\cdot\vec{A}\times\dot{\vec{A}}
\end{equation}
and from the third
\begin{equation}
{1\over 2M}P_+\left\{e\rlap
/{A}^\perp,\kappa_p\sigma_{\mu\nu}F^{\mu\nu}\right\}
P_+=\kappa_p{e^2\over M^2}\vec{S}\cdot\vec{A}\times\dot{\vec{A}}
\end{equation}
When added to the pole contributions the result can be expressed in the general
form\cite{53}
\begin{eqnarray}
\mbox{Amp}&=&\hat{\epsilon}\cdot\hat{\epsilon}'A_1+\hat{\epsilon}'\cdot
\vec{k}\hat{\epsilon}\cdot\vec{k}'A_2+i\vec{\sigma}\cdot(\hat{\epsilon}'
\times\hat{\epsilon})A_3\nonumber\\
&+&i\vec{\sigma}\cdot(\vec{k}'\times\vec{k})\hat{\epsilon}'\cdot\hat{\epsilon}
A_4+i\vec{\sigma}\cdot[(\hat{\epsilon}'\times\vec{k})\hat{\epsilon}\cdot\vec{k}'
-(\hat{\epsilon}\times\vec{k}')\hat{\epsilon}'\cdot\vec{k}]A_5\nonumber\\
&+&i\vec{\sigma}\cdot[(\hat{\epsilon}'\times\vec{k}')\hat{\epsilon}\cdot\vec{k}'
-(\hat{\epsilon}\times\vec{k})\hat{\epsilon}'\cdot\vec{k}]A_6\nonumber\\
\end{eqnarray}
with\footnote{Here we
have used the identity
$$\vec{\sigma}\cdot(\hat{\epsilon}'\times\vec{k}')\times(\hat{\epsilon}\times
\vec{k})=\vec{\sigma}\cdot(\vec{k}'\times\vec{k})\hat{\epsilon}\cdot
\hat{\epsilon}'
+\vec{\sigma}\cdot(\hat{\epsilon}'\times\hat{\epsilon})\vec{k}'\cdot\vec{k}
+\vec{\sigma}\cdot(\hat{\epsilon}\times\vec{k}')\hat{\epsilon}'\cdot\vec{k}
-\vec{\sigma}\cdot(\hat{\epsilon}'\times\vec{k})\hat{\epsilon}\cdot\vec{k}'
$$}
\begin{eqnarray}
A_1&=&-{e^2\over M},\quad A_2={1\over M^2\omega},\quad A_3={e^2\omega\over
2M^2}
(1+2\kappa-(1+\kappa)^2\hat{k}\cdot\hat{k}')\nonumber\\
A_4&=&-A_5=-{e^2(1+\kappa)^2\over 2M^2\omega},\quad A_6=-{e^2(1+\kappa)
\over 2M^2\omega}
\end{eqnarray}
which agrees with the usual result derived in this order via
Low's theorem.\cite{54}

\begin{figure}
\vspace{2.5in}

\caption{Pole diagrams for Compton scattering.}
\end{figure}

A full calculation at order $q^3$ must also, of course,
include loop contributions.  Using the lowest
order (one-derivative) pion-nucleon interactions
\begin{eqnarray}
{\cal L}_{\pi NN}&=&{g_A\over F_\pi}\bar{N}\tau^aS\cdot q N\nonumber\\
{\cal L}_{\pi\pi NN}&=&{1\over 4F_\pi^2}v\cdot (q_1+q_2)\epsilon^{abc}\bar{N}
\tau_cN\nonumber\\
{\cal L}_{\gamma\pi NN}&=&{ieg_A\over F_\pi}\epsilon^{a3b}\bar{N}\epsilon\cdot
S\tau_bN
\end{eqnarray}
these can be calculated using the diagrams shown in Figure 4.  Of course, from
Eq. 171 the propagator for the nucleon must have the form $1/iv\cdot k$ where
k is the off-shell momentum.  Thus, for example, the seagull diagram, Figure
4a, is of the form
\begin{equation}
\mbox{Amp}=4e^2({g_A\over F_\pi})^2\hat{\epsilon}\cdot\hat{\epsilon}'
\int{d^4k\over (2\pi)^4}{S\cdot kS\cdot k\over v\cdot k(k^2-m_\pi^2)
((k+q_1-q_2)^2-m_\pi^2)}
\end{equation}
Since there are no additional counterterms at this order $q^3$, the sum of loop
diagrams must be finite and yields, to lowest order in energy and after
considerable calculation
\begin{eqnarray}
A_1^{loop}&=&\xi({11\omega^2\over 24m_\pi}+{t\over 48m_\pi}),\quad A_2^{loop}=
\xi({1\over 24m_\pi}),\quad A_3^{loop}=\xi({\omega t\over \pi m_\pi^2}+
{\omega^3\over 3\pi m_\pi^2})\nonumber\\
A_4^{loop}&=&\xi({\omega\over 6\pi m_\pi^2}),\quad A_5^{loop}=-A_6^{loop}=
-\xi({13\omega\over 12\pi m_\pi^2}),
\end{eqnarray}
and $\xi=g_A^2/8\pi F_\pi^2.$

\begin{figure}
\vspace{2.5in}

\caption{Loop diagrams for Compton scattering.  Diagrams b,c,d must also
include cross terms.}
\end{figure}

The experimental implications of these results may be seen by first considering
the case of an unpolarized proton target.  Then writing
\begin{equation}
\mbox{Amp}_{unpol}=\{\hat{\epsilon}\cdot\hat{\epsilon}'(-{e^2\over M}+4\pi
\alpha_E\omega^2)+(\hat{\epsilon}\times\vec{k})\cdot(\hat{\epsilon}'\times
\vec{k}')4\pi\beta_M\}
\end{equation}
where $\alpha_E,\beta_M$ are the proton electric and magnetic polarizabilities,
we identify the one loop chiral predictions\cite{55}
\begin{equation}
\alpha_E^{theo}=10\beta_M^{theo}={5e^2g_A^2\over 384\pi^2F_\pi^2m_\pi}=
13.6\times 10^{-4}\mbox{fm}^3
\end{equation}
which are in reasonable agreement with the recently measured values\cite{56}
\begin{equation}
\alpha_E^{exp}=(10.4\pm 0.6)\times 10^{-4}\mbox{fm}^3,\qquad
\beta_M^{exp}=(3.8\mp 0.6)\times
10^{-4}\mbox{fm}^3
\end{equation}
For the case of spin-dependent forward scattering, we find
\begin{equation}
{1\over 4\pi}\mbox{Amp}=f_1(\omega^2)\hat{\epsilon}\cdot\hat{\epsilon}'
+i\omega f_2(\omega^2)\vec{\sigma}\cdot\hat{\epsilon}'\times\hat{\epsilon}
\end{equation}
Then we find
\begin{eqnarray}
f_1(\omega^2)&=&-{e^2\over 4\pi M}+(\alpha_E+\beta_M)\omega^2+{\cal
O}(\omega^4)
\nonumber\\
f_2(\omega^2)&=&-{e^2\kappa_p^2\over 8\pi^2M^2}+\gamma_S\omega^2+{\cal
O}(\omega^4)
\end{eqnarray}
where $\gamma_S$ is a sort of spin-polarizability.
Assuming that the amplitudes $f_1,f_2$ obey once-subtracted and unsubtracted
dispersion relations respectively we find the sum rules
\begin{eqnarray}
\alpha_E+\beta_M&=&{1\over 4\pi^2}\int_{\omega_0^\infty}{d\omega\over \omega^2}
(\sigma_+(\omega)+\sigma_-(\omega))\nonumber\\
{\pi e^2\kappa_p^2\over 2M^2}&=&\int_{\omega_0}^\infty{d\omega\over \omega}
[\sigma_+(\omega)-\sigma_-(\omega)]\nonumber\\
\gamma_S &=& {1\over 4\pi^2}\int_{\omega_0}^\infty{d\omega\over \omega^3}
[\sigma_+(\omega)-\sigma_-(\omega)]
\end{eqnarray}
where here $\sigma_\pm(\omega)$ denote the photoabsorption cross sections
for scattering cirvularly polarized photons on polarized nucleons for total
$\gamma N$ helicity 3/2 and 1/2 respectively.
Here the first is the well-known unitarity sum rule for the sum of the electric
and magnetic polarizabilities, while the second is the equally familiar
Drell-Hearn-Gerasimov sum rule.\cite{58}  The third is less well known, but
follows
from that of DHG and offeres a new check of the chiral predictions.

A second venue wherein chiral methods offer new predictive power is that
of threshhold photoproduction.  Here what is measured is the s-wave or
$E_{0+}$ multipole, defined via
\begin{equation}
\mbox{Amp}=4\pi(1+\mu)E_{0+}\vec{\sigma}\cdot\hat{\epsilon}+\ldots
\end{equation}
where $\mu=m_\pi/M$.  In the case of charged photoproduction, things are
relatively simple, as
the dominant contribution of the amplitude, occurring at one derivative
level is the so-called Kroll-Ruderman term given in Eq. 166.\cite{59}  In
addition, at
the two derivative level there exists a second contact term which arises
from
\begin{equation}
{\cal L}^{(2)}_{\pi\gamma NN}={eg_A\over 8MF_\pi}v\cdot qP_+[(1+\tau_3)
\rlap /{A}^\perp,\gamma_5\tau^a]P_+={eg_A\over 2MF_\pi}S\cdot\epsilon
v\cdot q(\tau^a+\delta^{a3})
\end{equation}
Adding these two contributions yields the result\cite{60}
\begin{eqnarray}
E_{0+}&=&\pm{1\over 4\pi(1+\mu)}{eg_A\over \sqrt{2}F_\pi}(1\mp{\mu\over 2})=
{eg_A\over 4\sqrt{2}F_\pi}\left(\begin{array}{cc}
1-{3\over 2}\mu & \pi^+ \\
-1+{1\over 2}\mu & \pi^-
\end{array}\right)\nonumber\\
&=&\left\{\begin{array}{ll}
+26.3 \times 10^{-3}/m_\pi  & \pi^+n \\
-31.3 \times 10^{-3}/m_\pi  & \pi^-p
\end{array}\right.
 \end{eqnarray}
The numerical predictions are found to be in excellent agreement with the
present experimental results,
\begin{equation}
E_{0+}^{exp}=\left\{\begin{array}{ll}
(+27.9\pm 0.5)\times 10^{-3}/m_\pi\cite{95} & \pi^+n \\
(+28.8\pm 0.7)\times 10^{-3}/m_\pi\cite{96} &   \\
(-31.4\pm 1.3)\times 10^{-3}/m_\pi\cite{95} & \pi^-p \\
(-32.2\pm 1.2)\times 10^{-3}/m_\pi\cite{97} &
\end{array}\right.
\end{equation}
However, these results are old emulsion measurements involving significant
extrapolation to threshold.  A new experiment is being run this summer
at Saskatoon which will explore the region only 1 MeV above threshold.

More challenging is the case of neutral photoproduction, for which the
one-derivative contribution vanishes.  In this case the leading contribution
arises from the two derivative term given in Eq. 171, augmented by the
three derivative contribution from the pole terms shown in Figure 5a.
The net result is
\begin{eqnarray}
\mbox{Amp}^{(2)}={eg_A\over 2F_\pi}\mu\hat{\epsilon}\cdot\vec{\sigma}\times
\left\{\begin{array}{cc}
1 & \pi^0p \\
0 & \pi^0n
\end{array}\right.
\end{eqnarray}
for the contact term and
\begin{equation}
\mbox{Amp}^{(3)}=-{e\over 2M}[S\cdot\epsilon,S\cdot k](1+\kappa_p)
{1\over v\cdot q}
{g_A\over 2MF_\pi}S\cdot(2p-q)m_\pi=
-{eg_A\over 4F_\pi}\mu^2(1+\kappa_p)\vec{\sigma}\cdot
\vec{\epsilon}
\end{equation}
for the pole terms (Note: only the cross term is nonvanishing at threshhold.)
Finally we must append the contribution of the loop contributions which arise
from the graphs shown in Figure 5b,c
\begin{equation}
\mbox{Amp}^{loop}=-{eg_AM\over 64\pi
F_\pi^2}\mu^2\vec{\sigma}\cdot\vec{\epsilon}
\end{equation}
The result is the prediction\cite{61}
\begin{equation}
E_{0+}={eg_A\over 8\pi M}\mu\{1-[{1\over 2}(3+\kappa_p)+({M\over 4F_\pi})^2]\mu
+{\cal O}(\mu^2)\}
\end{equation}
However, comparison with experiment is tricky because of the existence of
isotopic spin breaking in the pion and nucleon masses and the feature that
there exist two thresholds---one for $\pi^0p$ and the second for
$\pi^+n$---only
7 MeV apart.  When the physical masses of the pions are used the soon to be
published data from both Mainz and from Saskatoon are rumored to agree with the
chiral
prediction.\cite{62}

\begin{figure}
\vspace{2.5in}
\caption{Diagrams for neutral pion photoproduction.  Each should be
accompanied by an appropriate cross term.}
\end{figure}

There also exists a chiral symmetry prediction for the reaction $\gamma n
\rightarrow \pi^0n$
\begin{equation}
E_{0+}=-{eg_A\over 8\pi M}\mu^2\{{1\over 2}\kappa_n+({M\over 4F_\pi})^2\}
\end{equation}
However, the experimental measurement of such an amplitude involves
considerable challenge, and must be accomplished either by use of a deuterium
target with the difficult subtraction of the proton contribution and of
meson exchange contributions or by use of a polarized ${}^3$He target,
wherein one must account for the $\sim 10\%$ component of the wavefunction
which is not a simple polarized neutron.  Neither of these will be easy.

Other areas wherein chiral predictions can be confronted with experiment
include the electric dipole amplitude in electroproduction as well as the
P-wave multipoles in the ordinary photoproduction case.  One the challenges
which remains in this regard is the inclusion of the effects of the delta
resonance.  If one does this using a relativistic formalism, then the
power counting is no longer valid.  However, the problem of including the
delta in a heavy baryon formalism is not yet solved.

\section{Back to the Future}
We have spent a good deal of time now discussing the formalism of chiral
perturbation theory and I hope that I haven't given the impression
that this field is basically cut and dried and that there are few
remaining challenges,
for this certainly is {\it not} the case.  I shall close these lectures
then by outlining where I believe that there exists room for future work.
\subsection{Electroweak Goldstone Sector}
For the electroweak interactions of the pseudoscalar mesons the implications
of chiral symmetry have been well-developed by the work of Gasser, Leutwyler
and others.  In my view the remaining challenges are primarily experimental.
We have seen good agreement obtains in nearly all cases where chiral
predictions have been confronted with experimental tests, Nevertheless
there remain two possible problems.  One involves the discrepant values
for the charged pion polarizability discussed above, one of which disagrees
substantially from the value required by chiral invariance---
\begin{equation}
\bar{\alpha}_E^{\rm exp}=(6.8\pm 1.4)\times 10^{-4}\mbox{fm}^3\qquad {\rm vs.}
\qquad \bar{\alpha}_E^{\rm theo}=(2.8\pm 0.3)\times 10^{-4}\mbox{fm}^3
\end{equation}
Experiments to resolve this problem are proposed at DA$\Phi$NE,
at Fermilab, and at MAMI so we should have an answer before too long.
The other possible difficulty of which I am aware involves a probe of the
anomaly via $\gamma\rightarrow\pi^+\pi^-\pi^0$, where a disagreement exists
at the $3\sigma$ level\cite{31}
\begin{eqnarray}
{\rm Amp}(\gamma\rightarrow 3\pi)^{\rm exp}&=&12.9\pm 0.9\pm 0.5
\mbox{GeV}^{-3}\nonumber\\
&{\rm vs.}&\nonumber\\
{\rm Amp}(\gamma\rightarrow3\pi )^{\rm theo}&=&
9.7 \mbox{GeV}^{-3}
\end{eqnarray}
In this case there exists an approved experiment for the CLAS detector at
CEBAF.

\subsection{Nonleptonic Goldstone Sector}
Above we have explored the utility of chiral symmetry methods applied to
the semileptonic weak interactions of the Goldstone particles.  An
additional realm of Goldstone interactions opens if one considers the
arena of nonleptonic kaon decay, for which possible reactions are\cite{32}
\begin{itemize}
\item [i)] $K_S\rightarrow \pi\pi$
\item [ii)] $K_L\rightarrow \pi\pi\pi$
\item [iii)] $K_S\rightarrow \gamma\gamma$
\item [iv)] $K_L\rightarrow \pi^0\gamma\gamma$
\item [vi)] {\it etc.}
\end{itemize}
Although a great deal of work has been done on such processes, there remain
unanswered problems such as the origin of the $\Delta I={1\over 2}$ rule,
the discrepency between predicted and measured rates for the reaction
$K_L\rightarrow\pi^0\gamma\gamma$, quadratic and higher effects in the
$K_L\rightarrow\pi\pi\pi$ spectra, {\it etc.}.
\subsection{High Energy Extension}
As we have emphasized above, the strength of chiral perturbation theory
is that it gives predictions which are model-independent---deriving
solely from the symmetry properties of the underlying QCD
Lagrangian.  However, this carries with it a corresponding weakness in that
such predictions can only rigorously
be applied at energies small compared to the chiral scale of 1 GeV, while
experiments are not subject to such limitations.  Thus an important challenge
is to find ways by which to extend the validity of chiral methods to higher
energies.  This attempt has been made in a number of recent works, often
exploiting analyticity properties in the form of dispersion relations in
order to make this extension.  An example of such a program can be seen in
the analysis of the reaction $\gamma\gamma\rightarrow\pi^0\pi^0$ for which
there exist no tree-level contributions at either the two- or four-derivative
level but for which a finite one-loop prediction obtains.  Although this
one-loop prediction is {\it not} in agreement with recent experimental results
from
SLAC which extend up to about 1 GeV, the use of dispersion relations in order
to include the effects of loops to all orders has been shown to give very
good agreement over this
entire energy range.\cite{33}  However, this marriage between chiral
perturbative
and dispersive techniques is still new and it remains to be seen
whether it will be a lasting one.
\subsection{Calculating $L_i$ from Theory}
The original development of chiral perturbative techniques was phenomenological
is that values of the Gasser-Leutwyler counterterms were obtained purely
by empirical means.  A successful theory of everything (TOE)
would be able to
predict the size of such coefficients directly from the underlying QCD
Lagrangian, and this program remains an open challenge to future theoretical
work.  It is important to point out that important progress has
been made in this regard, and various techniques have been employed,
including
\begin{itemize}
\item [i)] Nambu-Jona-Lasinio models
\item [ii)] Lippman-Schwinger approach
\item [iii)] vector dominance
\item [iv)] lattice techniques
\end{itemize}
However, no approach is completely successful and much remains to be done.
\subsection{Extension to the Baryon Sector}
The success of application to the
processes $\gamma N\rightarrow\pi N$ and $\gamma N\rightarrow\gamma N$
is by no means clear---the convergence of the series may be too slow to
be of utility.  A marriage of heavy quark and dispersive methods
may be helpful here, but the verdict is still out.
Another important issue is inclusion of the $\Delta$ degrees of freedom.
It is clear that this must be done, as the $\Delta$ couples strongly and
its influence occurs even at near threshold energies.  The problem is
in order to do this in a consistent power counting scheme, a heavy
baryon expansion must be carried out for a spin 3/2 system, which is
challenging inasmuch as a Rarita-Schwinger spinor, used to describe such
a system, contains both spin 3/2 and two independent spin 1/2 degrees
of freedom.  Nevertheless, progress on this front has recently been reported
and such calculations should be forthcoming.\cite{70}

\section{Conclusions}
We have spent a great deal of time studying the consequences of
symmetry breaking in QCD.  We have learned that by exploiting
this breaking one can make rigorous contact between experimental
processes and the QCD Lagrangian which presumably underlies them.
It is interesting that of the three symmetry breaking mechanisms
which are possible in physics:
\begin{itemize}
\item[--] explicit: where the Lagrangian itself breaks the symmetry

\item[--] spontaneous: where the Lagrangian is symmetric but the ground state
is not

\item[--] anomalous: where the symmetry is broken as a result of quantization
\end{itemize}
{\it all three} are associated with ${\cal L}_{\rm QCD}$ and
by study of such symmetries and their breaking one can learn more about
both the symmetries and about QCD itself.

\end{document}